\documentclass[bibyear]{aa}  

\usepackage{graphicx}
\usepackage{txfonts}
\usepackage{float}
\usepackage{natbib}
\usepackage{hyperref} 
\hypersetup{colorlinks=true, linkcolor=blue, citecolor=blue, urlcolor=blue}
\usepackage{csvsimple}
\usepackage{arydshln} 
\usepackage{amsmath}

\usepackage{xcolor}
\newcommand{\teal}[1]{\textcolor{teal}{#1}}

\begin{document} 

\title{Atmospheric properties of AF\,Lep\,b with forward modeling}

\author{
P. Palma-Bifani\inst{1,2} \fnmsep\thanks{paulina.palma-bifani@oca.eu}, 
G. Chauvin\inst{1}, 
D. Borja\inst{1,3}, 
M. Bonnefoy\inst{3}, 
S. Petrus\inst{4}, 
D. Mesa\inst{5}, 
R. J. De Rosa\inst{6},  
R. Gratton\inst{5},\\
P. Baudoz\inst{2},
A. Boccaletti\inst{2},
B. Charnay\inst{2},
C. Desgrange\inst{3,7},
P. Tremblin\inst{8}, and
A. Vigan\inst{9}
}

\authorrunning{P. Palma-Bifani et al.}
\titlerunning{Atmospheric properties of AF\,Lep\,b}
\institute{
Laboratoire Lagrange, Université Cote d’Azur, CNRS, Observatoire de la Cote d’Azur, 06304 Nice, France 
\and LESIA, Observatoire de Paris, Univ PSL, CNRS, Sorbonne Univ, Univ de Paris, 5 place Jules Janssen, 92195 Meudon, France 
\and Université Grenoble Alpes, CNRS, IPAG, F-38000 Grenoble, France 
\and Núcleo Milenio Formación Planetaria - NPF, Universidad de Valparaíso, Av. Gran Bretaña 1111, Valparaíso, Chile 
\and INAF-Osservatorio Astronomico di Padova, Vicolo dell’Osservatorio 5, 35122 Padova, Italy 
\and European Southern Observatory, Alonso de Córdova 3107, Vitacura, Santiago, Chile 
\and Max Planck Institute for Astronomy, K\"onigstuhl 17, D-69117 Heidelberg, Germany 
\and Université Paris-Saclay, UVSQ, CNRS, CEA, Maison de la Simulation, F-91191, Gif-sur-Yvette, France 
\and Aix Marseille Université, CNRS, CNES, LAM, Marseille, France 
}

\date{Received \today}

\abstract{
About a year ago, a super-Jovian planet was directly imaged around the nearby young solar-type star AF\,Lep. The 2.8\,M$_{Jup}$ planet orbiting at a semi-major axis of 8.2\,au matches the predicted location based on the Hipparcos-Gaia astrometric acceleration.}
{We aim to expand the atmospheric exploration of AF\,Lep\,b by modeling all available observations obtained with SPHERE at VLT (between 0.95-1.65, at 2.105, and 2.253\,$\mu$m), and NIRC2 at Keck (at 3.8\,$\mu$m) with self-consistent atmospheric models.}
{To understand the physical properties of this exoplanet, we used \texttt{ForMoSA}. This forward-modeling code compares observations with grids of pre-computed synthetic atmospheric spectra using Bayesian inference methods. We used Exo-REM, an atmospheric radiative-convective equilibrium model, including the effects of non-equilibrium processes and clouds.
}
{From the atmospheric modeling we derive solutions at a low $T_{\rm eff}$ of $\sim$\,750\,K.
Our analysis also favors a metal-rich atmosphere ($>0.4$) and solar to super-solar carbon-to-oxygen ratio ($\sim$\,0.6). 
We tested the robustness of the estimated values for each parameter by cross-validating our models using the leave-one-out strategy, where all points are used iteratively as validation points. Our results indicate that the photometry point at 3.8\,$\mu$m strongly drives the metal-rich and super-solar carbon-to-oxygen solutions.
}
{Our atmospheric forward-modeling analysis strongly supports the planetary nature of AF\,Lep\,b. Its spectral energy distribution is consistent with that of a young, cold, early-T super-Jovian planet. We recover physically consistent solutions for the surface gravity and radius, which allows us to reconcile atmospheric forward modeling with evolutionary models, in agreement with the previously published complementary analysis done by retrievals. Finally, we identified that future data at longer wavelengths are mandatory before concluding about the metal-rich nature of AF\,Lep\,b.
}

\keywords{Planets and satellites: gaseous planets, atmospheres, composition, formation}

\maketitle

\section{Introduction}\label{S.intro}

\begin{table*}[!h]
\centering
\small
\caption{Description of the observations.}
\resizebox{\textwidth}{!}{%
\begin{tabular}{| l : lllllllll|}
\hline
UT Date & Tel./Instrument & Filters & Obs. strategy & Airmass & Seeing (") & $\tau_0$ (ms) & Field Rot. (deg) & Tot. Exp. (s) & Reference  \\ 
\hline\hline
2021-12-21 & Keck/NIRC2 & Lp        & ASDI & - & 0.5    & -  & 35.8 & 2376 & \cite{Franson2023AstrometricLep} \\ 
2022-10-16 & VLT/SPHERE & YJH, K1K2 & ASDI & 1.1/1.2 & 0.55 & 5.8 & 13.3 & 3584 & \cite{Mesa2023AFData} \\
2022-10-20 & VLT/SPHERE & YJH, K1K2 & RDI & 1.4/1.8 & 0.46 & 6.7 & 4.5  & 2688 & \cite{DeRosa2023DirectLeporis} \\
2022-12-20 & VLT/SPHERE & YJH, K1K2 & ASDI & 1.0/1.1 &  1.05 & 4.0 & 51.8 & 3584 & \cite{Mesa2023AFData}  \\
2023-02-03 & Keck/NIRC2 & Lp & ASDI & - & 0.6  & - & 85.5    & 7884 & \cite{Franson2023AstrometricLep}  \\
\hline
\end{tabular}}
\tablefoot{Table summarizing the observing conditions for all available datasets of AF\,Lep\,b.}
\label{table:obslog}
\end{table*}

\begin{figure*}[t!]
\centering
\includegraphics[width=0.95\hsize]{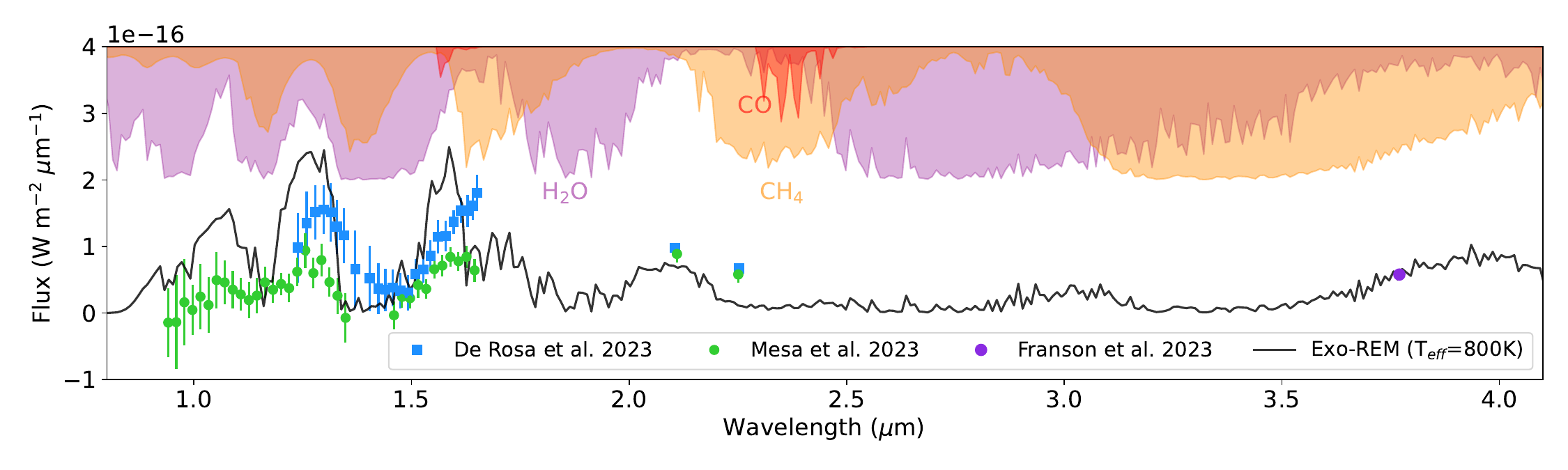}
\caption{Spectroscopic and photometric measurements recovered for AF\,Lep\,b from the original publications \citep{Mesa2023AFData, DeRosa2023DirectLeporis,Franson2023AstrometricLep}. 
The absorption features for H$_2$O, CO, and CH$_4$ are colored, which were computed by Exo-REM \citep{Charnay2021FormationK2-18b} at $T_{\rm eff}=800$\,K, log($g$)\,=\,4, [M/H]\,=\,1, and C/O\,=\,0.5. The black solid line represents the total flux for that model. 
}
\label{fig:Obs+HR8799e}
\end{figure*}

The era of exoplanet characterization started two decades ago with the atmospheric studies of hot and highly irradiated Jupiters using transmission and emission spectroscopy \citep{Charbonneau2002DetectionAtmosphere}. Such observations have been reported for more than 30 exoplanets, including mature (0.5-10 Gyr) hot Jupiters, hot Neptunes, and even super-Earths, and have allowed us to access information regarding the composition, spatial structures, and dynamics of exoplanetary atmospheres \citep{Guillot2022GiantInside-Out}. For young (1-100 Myr) exoplanets that can be spatially resolved, high-contrast imaging techniques (i.e., extreme adaptive optics, coronagraphy, and integral-field spectrographs) currently provide high-fidelity spectra at predominantly low-resolution, consisting of tens to thousands data points over a wide range of wavelengths (0.5-5\,$\mu$m), and can be acquired in a few telescope hours. These observations have allowed the community to explore the physical properties, composition, and clouds of a few young super-Jovians exoplanets \citep{Currie2022DirectPlanets}. 

For more than a decade, direct imaging surveys have targeted a broad sample of young, nearby stars to search for giant exoplanets beyond typically 10\,au with a low discovery rate. The recent results coupling \texttt{Gaia DR3} and \texttt{Hipparcos} proper motion promise to be an efficient target selection tool to narrow down the sample. The proper motion anomaly (PMa) can point towards the presence of companions, potentially planetary-mass ones \citep{Kervella2022StellarEDR3, Brandt2021Htof:Missions}, since they gravitationally impact the proper motion of their host star. AF\,Lep\,b is one of these first astrometrically-tagged exoplanets confirmed by high-contrast imaging by three different groups nearly simultaneously: \citet{Mesa2023AFData}, \citet{DeRosa2023DirectLeporis}, and \citet{Franson2023AstrometricLep}.

AF\,Lep\,A, also known as HD\,35850, HR\,1817, and HIP\,25486, is a 1.09\,$\pm$\, 0.06\,M$_{\odot}$ star of spectral type F8, and slightly sub solar metallicity ([Fe/H] $=-0.27\pm0.31$), located at 26.825\,$\pm$\,0.014\,pc (values updated in \citet{Zhang2023ELementalAtmospheres}). 
The age of the system is $24\pm3$\,Myr \citep{Bell2015ANeighbourhood}, measured from an absolute isochronal age scale considering its membership to the $\beta$\,Pictoris moving group. 
In \citet{Zhang2023ELementalAtmospheres} the authors revisit the atmospheric and orbital properties of AF\,Lep\,b updating the mass to 2.8\,$\pm$\,0.5\,M$_{\rm Jup}$ and the semi-major axis to 8.2\,$\pm1.5$\,au, using Hipparcos-Gaia DR3 PMa and relative astrometry. Their orbital fit is more accurate since they combined the information from the previous three works.
The estimated spectral type for this companion is late L ($>$L6) with a corresponding effective temperature ($T_{\rm eff}$) first estimated between 1000 and 1700\,K by \citet{Mesa2023AFData}. \citet{Zhang2023ELementalAtmospheres} updated the $T_{\rm eff}$ to be $\sim 800$\,K and constrained additional parameters as the surface gravity (log($g$)) to be 3.7\,dex and a potential metal enrichment on the atmosphere through a retrieval analysis.
In addition, the system likely hosts an unresolved debris disk at 46$\pm$9\,au, as shown in \citet{Pearce2022PlanetDiscs} by modeling the red excess of the spectral energy distribution (SED) of AF\,Lep\,A. In \citet{Pearce2022PlanetDiscs} they discuss that one additional planetary mass companion should be present between the orbit of AF\,Lep\,b and the inner edge of the debris disk to truncate the disk. However, the existence and location of the disk and planet c remain to be confirmed. 
 
The intriguing combination of AF\,Lep\,b's architecture and its unique position on the color-magnitude diagrams (L-T transition), where clouds and disequilibrium chemistry play a prominent role, makes it an exceptional laboratory for gaining insights into the formation and evolution of young planetary systems. 
Here, we study the atmosphere of AF\,Lep\,b based on forward modeling, complementary to the retrieval atmospheric study by \citet{Zhang2023ELementalAtmospheres}.
We present an overview of the data in Section \ref{S.Data}. We describe the atmospheric models in Section \ref{S.Atmo} and explore physically consistent solutions in Section \ref{S.loo}. We discuss the results and limitations in Section \ref{S.Discu} and our main takeaway points are in Section \ref{S.Concl}. A detailed view of the model setup and results can be found in Appendix \ref{app.models} and \ref{app.loo}.

\section{Available observations}\label{S.Data}

\subsection{Published data}

The currently available datasets are from the three publications reporting the discovery of AF\,Lep\,b.
The observing conditions and methodology for the three VLT/SPHERE epochs (one from \citet{DeRosa2023DirectLeporis} and two from \citet{Mesa2023AFData}) and the two Keck/NIRC2 epochs \citep{Franson2023AstrometricLep} are summarized in Table \ref{table:obslog}.
Using the initially published spectroscopic and photometric data points, we reconstructed the SED of AF\,Lep\,b as shown in Figure \ref{fig:Obs+HR8799e}. 
\citet{Mesa2023AFData} published observations from two epochs using SPHERE at the VLT at low-resolution (R$_{\lambda}\sim 30$) for the $YJH$ band and photometry points for the $K1$ and $K2$ filters using Angular and Spectral Differential Imaging (ASDI) as their observational technique (see Table \ref{table:obslog}). \citet{DeRosa2023DirectLeporis} observed the target using SPHERE at the VLT with the same filters and a similar spectral resolution but only one epoch. Their observational technique was Reference-star Differential Imaging (RDI).
\citet{Franson2023AstrometricLep} observed AF\,Lep\,b on two epochs at Keck using the NIRC2 vortex coronagraph in the $L'$ band filter. 
We constructed two sets of observations (dataset DR-F and M-F), corresponding to the data published by \citet{DeRosa2023DirectLeporis} and \citet{Mesa2023AFData}, respectively, combined with the photometric point from \citet{Franson2023AstrometricLep}. 
The K-band flux difference between the K1 and K2 points suggests the presence of potentially CH$_4$ on the atmosphere of AF\,Lep\,b, as visible in Figure \ref{fig:Obs+HR8799e} from the slope of the absorption feature. The absorption features are calculated by considering the cross-section of the specified molecule together with the collision-induced absorption (CIA) of H$_2$ – H$_2$ and H$_2$ – He and scattering within Exo-REM \citep{baudino2015interpreting}. We normalized the absorptions by the CIAs and applied an offset for visualization in Figure \ref{fig:Obs+HR8799e}.

\subsection{Flux discrepancies}

Both SPHERE datasets do not have the same wavelength coverage because, in \citet{DeRosa2023DirectLeporis}, only upper limits were provided for points below 1.22\,$\mu$m, which we did not include here.
We separated the datasets given that their flux level differs by more than 1\,$\sigma$ between 1.24 and 1.65\,$\mu$m (see Figure \ref{fig:Obs+HR8799e}). 
As shown in Table\,\ref{table:obslog}, all SPHERE observations were completed in relatively good conditions in terms of coherence time ($>4$\,ms) and exposure time on the target. 
Therefore, the flux difference could come from the relative high airmass sequence obtained by \citet{DeRosa2023DirectLeporis} in RDI that could be more sensitive to the temporal residuals of the extreme adaptive optics correction and of the non-common path aberrations. 
Another explanation could come from the data processing itself and the use of two different pipelines (\texttt{pyKLIP} from \citet{Pueyo2016DETECTIONMODELING} versus \texttt{Specal} from \citet{Galicher2018AstrometricSpeCal}). Since the K1 and K2 fluxes look consistent, it is more likely that the difference arises from the data reduction process of the IFS observations. 
AF\,Lep\,b is a faint source in the speckle-limited regime at a close separation that can challenge the available data reduction tools. 

\section{Atmospheric forward-modeling}\label{S.Atmo}

We proceed with forward modeling for the atmospheric study using a self-consistent pre-computed grid. This provides a gain in computation time and a complementary approach with respect to retrievals.
We implemented the forward modeling Python package called \texttt{ForMoSA}\footnote{\texttt{ForMoSA} documentation at \url{https://formosa.readthedocs.io}.}, which we have been developing over the past years. 

\begin{figure*}[ht!]
\centering
\includegraphics[width=\hsize]{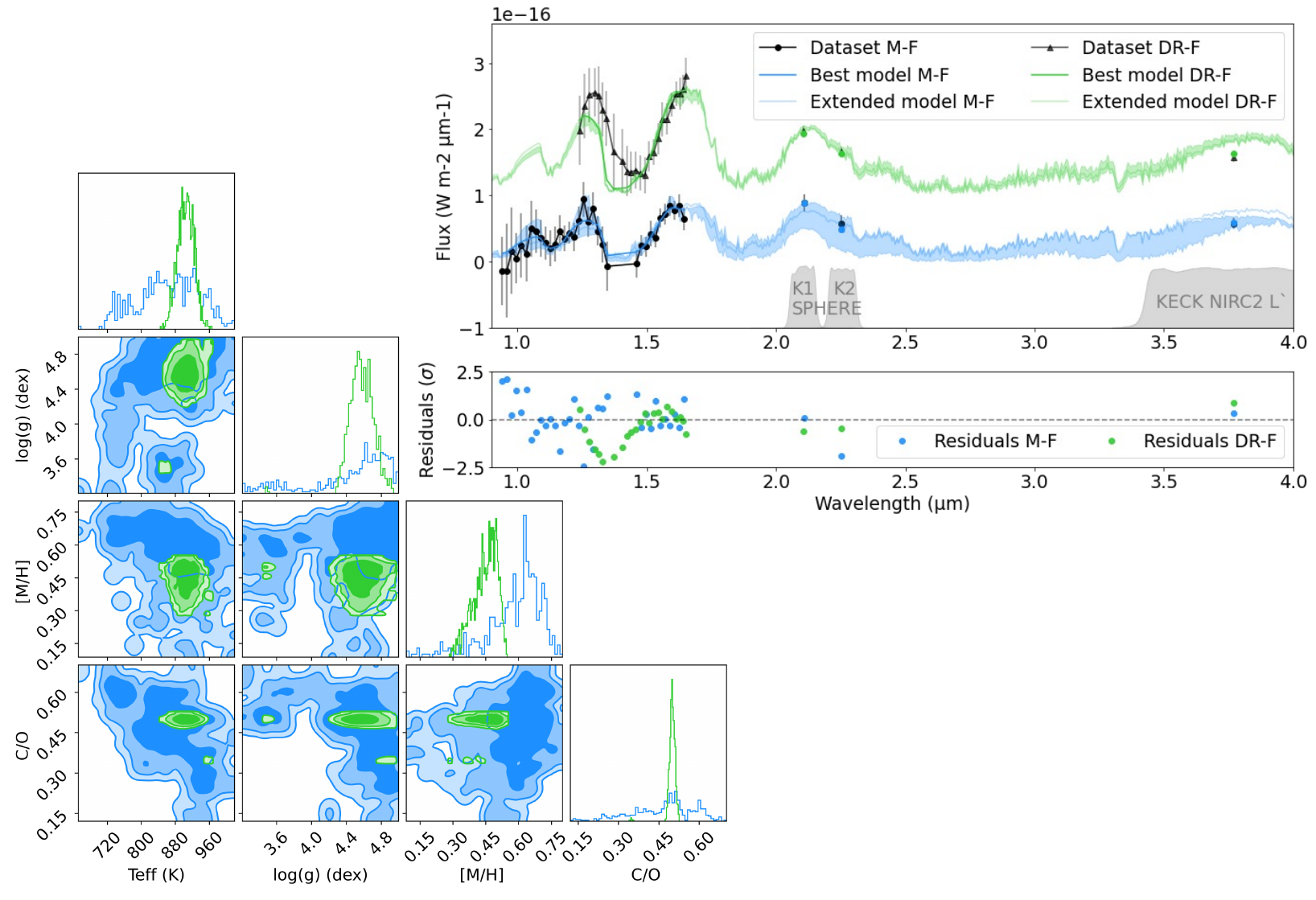}
    \caption{Best fit for both datasets using Exo-REM with only grid parameters. The left panel shows the posterior distributions.
    The upper right panel shows the models, with an offset applied to the DR-F dataset. 
    We extrapolated the models to visualize the complete spectral energy distribution from 0.9 to 4\,$\mu$m at R$_{\lambda}$\,=\,100. The colored areas represent rough variations on the models when considering 1\,$\sigma$ uncertainties for each parameter.}
\label{fig:FitExo}
\end{figure*}

\subsection{\texttt{ForMoSA}}\label{SS.formosa}

\texttt{ForMoSA} is a Bayesian Forward modeling code based on a nested sampling algorithm performing parameter exploration given a likelihood function \citep{Skilling2006Skilling833-860}. A full description of the code and applications on planetary-mass companions can be found in \citet{Petrus2021Medium-resolutionB} for HIP\,65426\,b, in \citet{Petrus2023X-SHYNE:Analogs} for VHS\,1256\,AB\,b, and in \citet{Palma-Bifani2023PeeringPic} for AB\,Pic\,b.
Here, we exploit its capabilities for studying the atmosphere of a companion at low spectral resolution (R$_{\lambda}\sim 30$).
Several grids with different atmospheric physical descriptions are currently available in the community.

For this work, we used Exo-REM, an atmospheric radiative-convective equilibrium model developed and presented in \citet{baudino2015interpreting}, \citet{charnay2018self, Charnay2021FormationK2-18b}, and \citet{refId0}. In this model, the atmosphere is split into pressure levels where the flux is calculated iteratively, assuming radiative-convective equilibrium and initial abundances from \citet{Lodders2010ExoplanetChemistry}. 
To calculate the opacities, this model considers the CIA of H$_2$ – H$_2$ and H$_2$ – He, ro-vibrational bands from nine molecules (H$_2$O, CH$_4$, CO, CO$_2$, NH$_3$, PH$_3$, TiO, VO, and FeH), and resonant lines from Na and K. In addition, Exo-REM includes non-equilibrium chemistry between C-, O-, and N-bearing compounds due to vertical mixing, parametrized by the eddy mixing coefficient as described in Section 2.2.2 from \citet{charnay2018self}.
The cloud description of the models is specially defined to suit the L/T transition, including iron and silicate clouds.
For the grid implemented here, the free parameters are $T_{\rm eff}$ from 400 to 2000\,K, log($g$) from 3 to 5\,dex, [M/H] from -0.5 to 1.0, and C/O from 0.1 to 0.8, at a spectral resolution of $\sim$\,10\,000 between 0.7 and 251\,$\mu$m. In Exo-REM [M/H] refers to the bulk metallicity, meaning the metal abundance with respect to hydrogen relative to the solar value, where a metal is every element heavier than helium.

In addition to the parametrization of the grids, we explored solutions for the radius ($R$) and interstellar medium extinction ($A_v$).
By default, the re-normalization of the synthetic spectrum is done analytically in \texttt{ForMoSA}. However, when the distance ($d$) to the target is known, we can include $R$ as a free parameter and re-normalize by $(R / d)^{2}$.
We explored three different combinations of free parameters using uniform priors between the ranges listed in Table \ref{tab:ForMoSA}.
We label the results corresponding to the parameter setup described below: 
\begin{enumerate}
    \item grid parameters only, 
    \item grid including $R$ and fixing the distance to AF\,Lep, 
    \item we added $A_v$ to the setup defined in 2.
\end{enumerate}

More details regarding the priors and the different setups can be found in the Appendix \ref{app.models}.

\subsection{Results for datasets DR-F and M-F}\label{SS.results} 

\renewcommand{\arraystretch}{1.2}
\begin{table}[ht]
\centering
\small
\caption{\texttt{ForMoSA} initial results for both datasets.}
\begin{tabular}{| l l : c c |} 
\hline
Model & $Parameter$ & Dataset DR-F & Dataset M-F \\ 
\hline\hline
E1 & $T_{\rm eff} \ (K)$ & $900 \pm 20$ & $830 \pm 90$ \\
& log($g$)  & $4.4 \pm 0.1$ & $> 4.2$ \\
& $[M/H]$ & $0.5 \pm 0.1$ & $0.7 \pm 0.1$ \\
& C/O & $0.5 \pm 0.01$ & $0.5 \pm 0.1$ \\
\hline\hline
E2 & $R$ ($R_{\rm Jup}$) & $0.9\pm0.1$ & $0.9\pm0.2$ \\
& log($L$/$L\mathrm{_{\odot}}$) & $-5.35\pm0.03$ & $-5.4\pm0.1$ \\
\hline\hline
E3 & $A_v$ & $7.2_{-1.3}^{+1.0}$ & $0.5_{-0.5}^{+1.0}$\\
\hline
\end{tabular}
\label{tab:best}
\tablefoot{The detailed values for each model run are listed in Table \ref{tab:ForMoSA}}
\end{table}

A convenient feature of nested sampling is that it returns the natural logarithm of the Bayesian evidence (log(z)) together with a measure of the information (h). 
The information is defined in \citet{Skilling2006Skilling833-860} as the logarithm of the fraction of the prior mass that contains the bulk of the posterior mass, also refer to as the "peakiness" of the likelihood function in Nestle\footnote{Nestle documentation at \url{http://kylebarbary.com/nestle/}}, the nested sampling algorithm implemented in \texttt{ForMoSA}.  
A log(z) in \texttt{ForMoSA} will always be negative, meaning that the closer to 0, the better the fit. If h is zero, the likelihood function is constant and does not carry any information. The parameter h is used to compute the statistical sampling uncertainty on log(z) as
\begin{equation}
    \rm log(z)_{ \rm err} = \sqrt{\rm h / n}  \, ,
\end{equation}
with n representing the number of living points.

The derived physical properties for each run are listed in Table\,\ref{tab:ForMoSA}.
The log(z) has higher values for both datasets when fitting only for the atmospheric model grid parameters. 
Therefore, we cannot constrain the presence of additional components as interstellar dust causing extinction (A$_{v}$) or any other additional parameter implemented in \texttt{ForMoSA}. 

We show the best-modeled spectra and the posterior parameter distributions in Figure \ref{fig:FitExo} for both datasets. 
The errors given by \texttt{ForMoSA} represent 1$\sigma$ ranges assuming asymmetric Gaussian distributions of the posteriors. They result from the propagation of the data errors through the Bayesian inversion, meaning that they do not account for systematic deviations of the model versus the data and should be treated as purely statistical. 
A significant result is that ExoREM favors the tail of the coldest solutions initially found by \citet{Mesa2023AFData}. 
The best-fit show that AF\,Lep\,b has a $T_{\rm eff}$ of $900\pm20$\,K and $830\pm90$\,K for the DR-F and M-F datasets modeled with Exo-REM respectively (see Table \ref{tab:best}).

The log($g$) best solution for both datasets is high ($> 4.2$\,dex). In addition, the derived $R$ is too small and nonphysical ($\sim$\,0.9\,R$_{\rm Jup}$) as often derived with atmospheric models (often referred to as the small radii problem).  
Regarding the composition, we have some preliminary constraints on [M/H] and C/O ratio, which must be considered cautiously. The [M/H] is $>$0.4 for both datasets, consistent with super-solar values. 
For the C/O ratio, the solutions range between 0.4 and 0.8, compatible with a solar to super-solar value (C/O$_{\odot}$\,$=$\,0.55; \citet{Asplund2009TheSun}). 
The bolometric luminosity (log($L$/$L_\odot$)) derived by \texttt{ForMoSA} from the models without $A_v$ is $\sim -$5.4 for both datasets, fainter than the estimations of \citet{DeRosa2023DirectLeporis}.

The $A_v$ exploration differs considerably for both datasets.
For the M-F dataset, $A_v$ is $\sim$\,0.5\,mag, which could be caused by a circumplanetary disk (CPD). The detection of emission lines on the companions 2MASS\,J0249-0557\,c \citep{Chinchilla2021StrongC} from the $\beta$ Pictoris moving group suggests that CPDs might surround exoplanets such as AF\,Lep\,b.
For DR-F, the $A_v$ value is unexpectedly high ($\sim$\,7\,mag). Such high values are expected for a young planet such as PDS\,70\,c \citep{Wang2021Constrainingsup/sup}, but rather unexpected for older systems such as AF\,Lep. We further discuss this high $A_v$ result in Section \ref{ss.dr}.

\section{Exploring physically consistent solutions}\label{S.loo}

\subsection{Evolutionary models}\label{SS.results_evo} 

\begin{figure}[t!]
\centering
\includegraphics[width=\hsize]{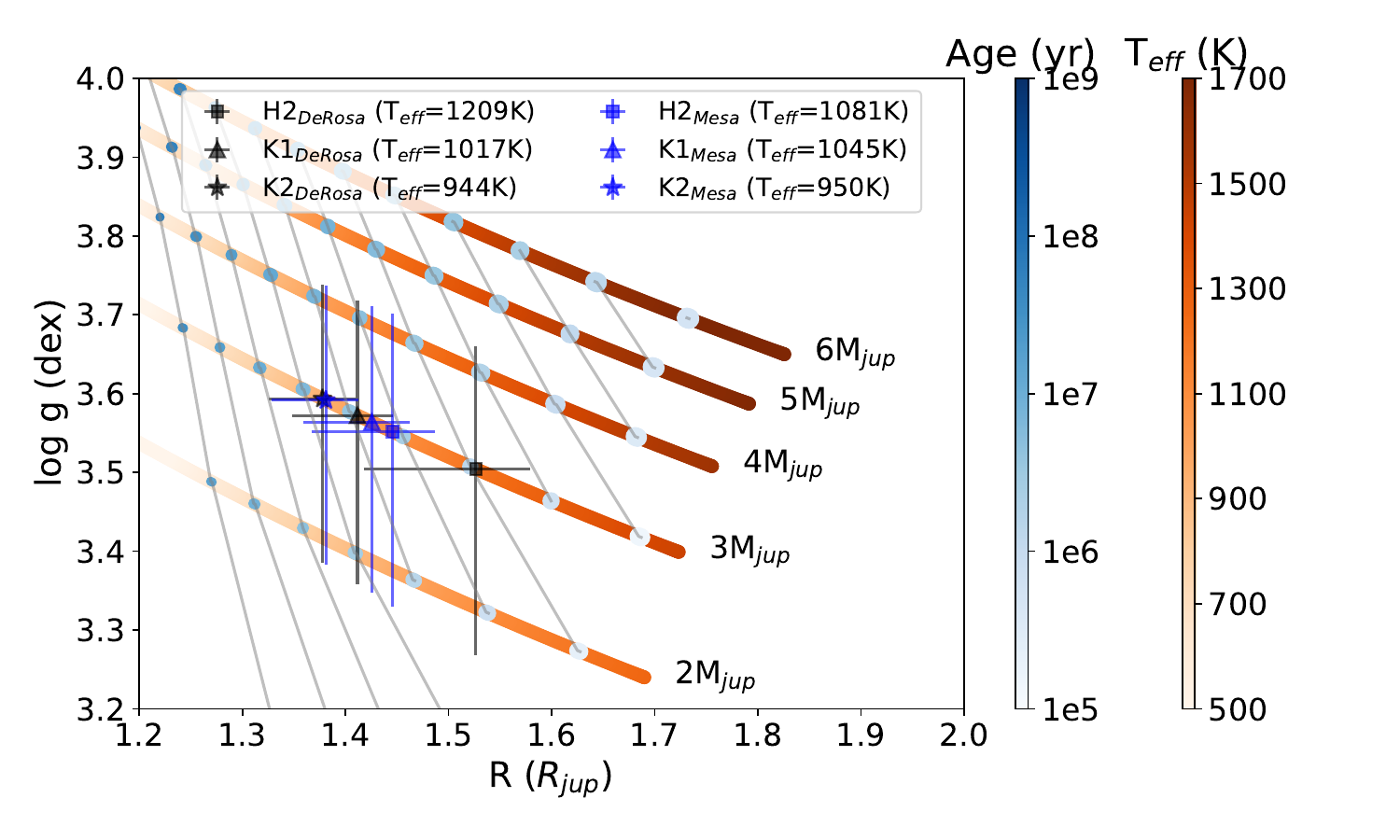}
\caption{Evolutionary models from BEX-Hot-COND\,03. The model curves are a function of $R$ and log($g$), which are represented in terms of their age and $T_{\rm eff}$ variations by the use of two different color scales. The reddish color map lines represent iso-mass curves. The gray lines show iso-temperature curves where the blueish scale marks the age variations.
We predict for AF\,Lep\,b a $T_{\rm eff}$ ranging from 950 to 1200\,K, R between 1.2 and 1.4\,R$_{\rm Jup}$, and log($g$) $\sim$ 3.8\,dex using the absolute magnitudes from the photometry points published by \citet{DeRosa2023DirectLeporis} (black) and \citet{Mesa2023AFData}) (blue). The errors from the magnitude values were propagated onto log($g$) and $R$.}
\label{fig:evolMod}
\end{figure}

In section \ref{SS.results} we had a preliminary look over the physical properties of this companion and identified high log($g$) and small $R$ solutions for both datasets. Evolutionary models provide an essential extra piece of information to estimate and corroborate parameters such as log($g$), radius, and luminosity. 
We considered the BEX-Hot-COND\,03 models \citep{Marleau2019ExploringB}, which tabulate how the flux varies for different spectral bands as a function of the age, radius, luminosity, $T_{\rm eff}$, and log($g$). 
We report the predictions of these models in terms of iso-mass (reddish lines) and iso-temperature (gray lines) curves as a function of $R$ and log($g$) for different ages (see Figure \ref{fig:evolMod}), as done previously in \citet{Palma-Bifani2023PeeringPic}. A way of interpreting the curves in Figure \ref{fig:evolMod} is by observing that a planet with a fixed mass contracts over time, leading to an increase in log($g$). 

We extracted from the tabulated variations the predicted log($g$) and $R$ values for AF\,Lep\,b, by using the reported photometry in the $H2$, $K1$, and $K2$ filters from \citet{DeRosa2023DirectLeporis} and \citet{Mesa2023AFData}, together with the age as input. To avoid interpolations, we used an approximated mass of 3\,$\pm1$\,M$_{jup}$ for AF\,Lep\,b.
Note that the $H2$ data point predictions from \citet{DeRosa2023DirectLeporis} are slightly off, indicating that their spectrum flux level and/or wavelength calibration need to be revisited.
Apart from that point, we observe that all other points agree with a lower log($g$) of 3.85$\pm$0.15\,dex, and a larger R of 1.3\,$\pm$\,0.1\,R$_{\rm Jup}$ (see Figure \ref{fig:evolMod}). Similar inconsistencies between evolutionary models and forward modeling have already been reported, as for the mid-L-type exoplanet HIP\,65426\,b \citep{Carter2023Them}. This degeneracy of the atmospheric models seems to resolve when including observations at longer wavelengths for planets at the L-T transition, as explored with JWST/MIRI observations by \citet{Boccaletti2023ImagingCoronagraph}.

\subsection{DR-F dataset issues}\label{ss.dr}

Several issues regarding the DR-F dataset can be identified from the previous sections.
In Figure \ref{fig:FitExo}, we observe that the M-F dataset matches better to the atmospheric models than the DR-F dataset. The main spectral features are red-shifted for the \citet{DeRosa2023DirectLeporis} spectrum with respect to the best model.
In the corner plot in Figure \ref{fig:FitExo}, we also observe that the DR-F posteriors are much tighter than M-F, but as mentioned before, the errors provided by \texttt{ForMoSA} are purely statistical. Therefore, the tighter posteriors do not mean that the DR-F dataset provides tighter constraints and is rather related to the shifted features observable in Figure \ref{fig:FitExo} between 1.3 and 1.4\,$\mu$m together with the different spectral coverage of both datasets.
In addition, the high $A_v$ result ($\sim$\,7\,mag) from model DR-F E3 with a corresponding $T_{\rm eff}$ of $\sim$\,1870\,K is somewhat unrealistic, and is telling us that the models do not well reproduce the flux level of the DR-F observations.
Finally, the $H2$ data point from \citet{DeRosa2023DirectLeporis} evolutionary models prediction is off too.
The overall impact of these discrepancies remains minimal in \texttt{ForMoSA} with ExoREM given that the physical properties are consistent within 3\,$\sigma$ (see Table \ref{tab:best} and right panel of Figure \ref{fig:FitExo}). 
However, to avoid propagating the wavelength or flux calibration issue onto our solutions, we only use the M-F dataset in the following sections.

\subsection{Prior in log($g$) and final results}\label{SS.finalresults} 

\renewcommand{\arraystretch}{1.6}
\begin{table}[ht]
\centering
\small
\caption{Results with and without restrictive log($g$) prior for M-F.}
\begin{tabular}{| c : c c |} 
\hline
$Parameter$ & Default prior & Restrictive prior \\ 
\hline\hline
$T_{\rm eff} \ (K)$ & $854.0^{+63.0}_{-81.0}$ & $750.0^{+99.0}_{-36.0}$ \\
log($g$)  & $4.97^{+0.17}_{-0.34}$ & $3.68^{+0.07}_{-0.06}$ \\
$[M/H]$ & $0.69^{+0.04}_{-0.06}$ & $0.6^{+0.08}_{-0.13}$ \\
C/O & $0.49^{+0.08}_{-0.12}$ & $0.61^{+0.05}_{-0.09}$ \\
R (R$_{\rm Jup}$) & $0.84^{+0.23}_{-0.11}$ & $1.23^{+0.13}_{-0.18}$ \\
log($L$/$L\mathrm{_{\odot}}$) & $-5.43\pm0.05$ & $-5.36\pm0.05$ \\
\hline\hline
log(z) & $-16.7\pm0.2$  & $-16.4\pm0.2$\\
\hline
\end{tabular}
\label{tab:finalbest}
\end{table}

\begin{figure}[ht!]
\centering
\includegraphics[width=\hsize]{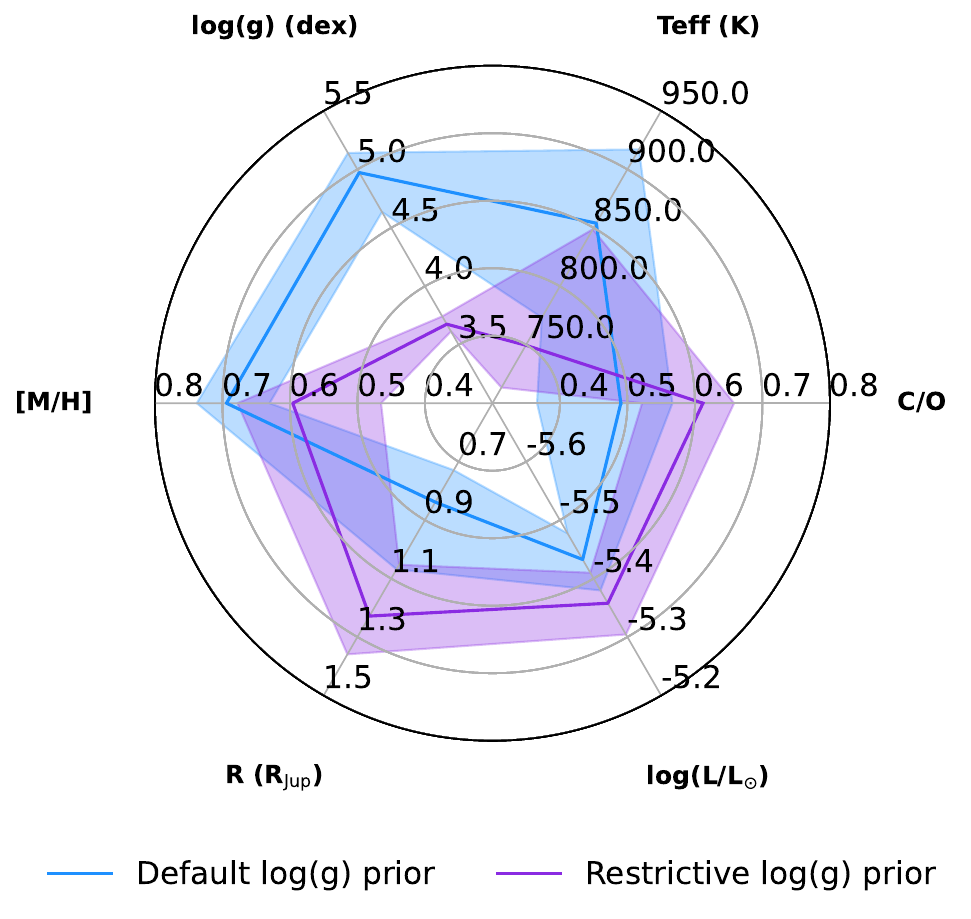}
\caption{Comparison of the results with and without the restrictive log($g$) prior. The colored areas represent 1\,$\sigma$ uncertainties, assuming Gaussian distributions of the posteriors.}
\label{fig:radar}
\end{figure}

\begin{figure*}[ht!]
\centering
\includegraphics[width=\hsize]{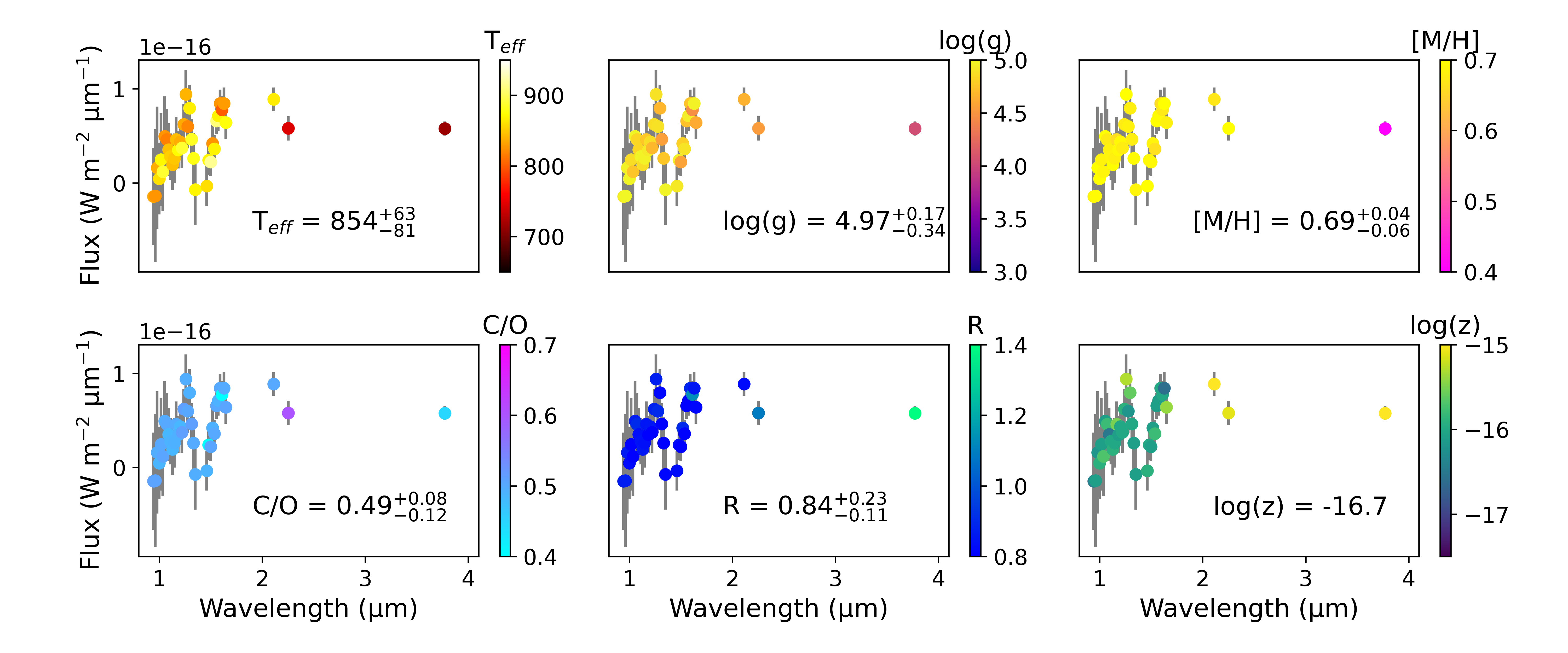}
    \caption{Leave-one-out analysis for the model with default priors.
    This figure consists of 6 panels where we plot the M-F dataset and analyze the behavior of the posteriors for each parameter. We assign a color level to each point of the spectrum representing the derived values when that point was kept out of the fit. In each panel, we wrote the posterior value for the case where all points were included to guide the comparison.
    }
\label{fig:loo0}
\end{figure*}

\begin{figure*}[ht!]
\centering
\includegraphics[width=\hsize]{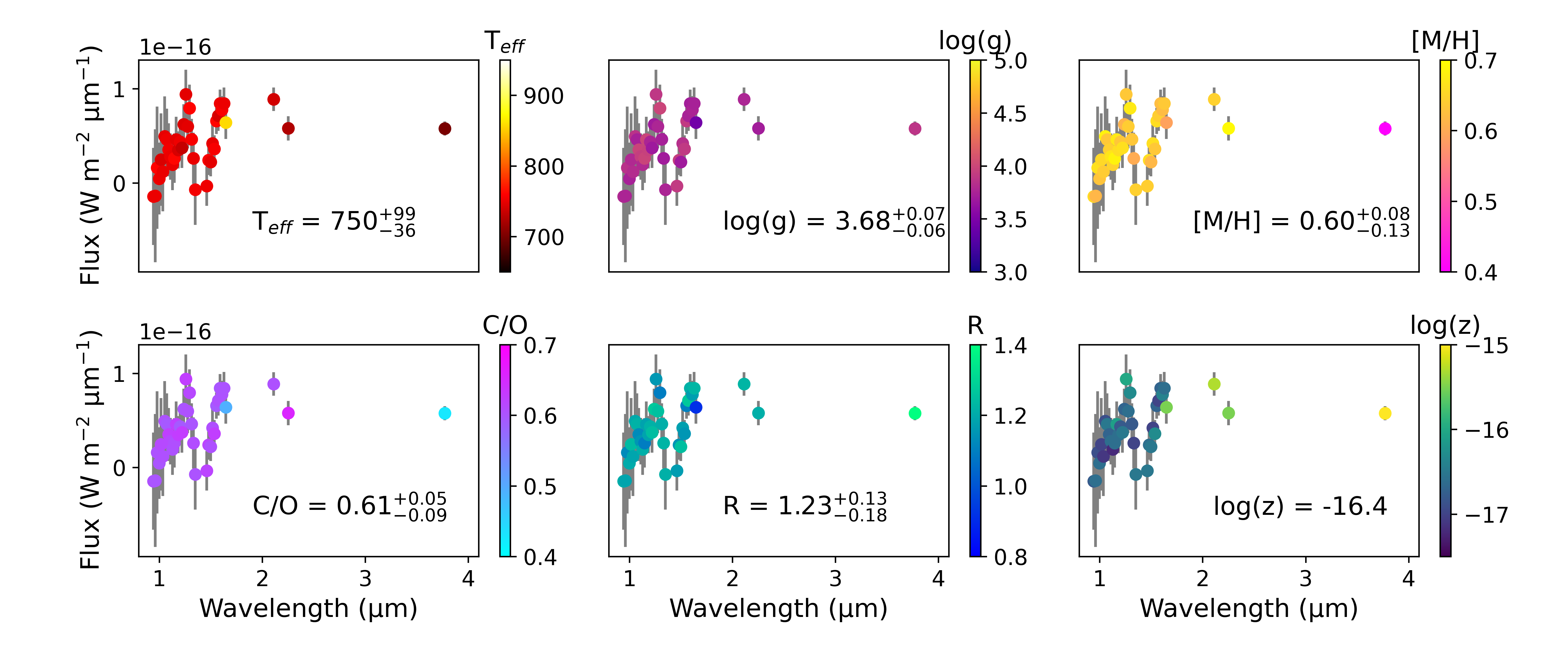}
    \caption{Same leave-one-out analysis as in Figure \ref{fig:loo0} but for the models with restrictive log($g$) prior.}
\label{fig:loo1}
\end{figure*}

When looking carefully at the log($g$) posterior distribution in the corner plot in Figure \ref{fig:FitExo} we observe a double-peaked solution. The primary solution peaks at 4.8\,dex, and a second family of solutions peaks at log($g$) around 3.6\,dex, albeit with a lower probability.
This lower log($g$) solution is consistent with the evolutionary models' predictions and can be recovered when taking the mass value of 2.8\,$\pm$\,0.6\,M$_{jup}$ derived by the orbital modeling in \citet{Zhang2023ELementalAtmospheres} and the estimated radius from evolutionary models from above into the following equation
\begin{equation}
    log(g) = log(\frac{M G}{R^2}),
\end{equation}
where $G$ is the gravitational constant. When replacing the numbers and propagating the errors, log($g$) is expected to be 3.7\,$\pm$\,0.1\,dex, in agreement with the lower intensity peak of the posterior distribution and the retrievals solutions by \citet{Zhang2023ELementalAtmospheres}.

To reconcile the evolutionary models' predictions with the atmospheric models, we add a restrictive uniform prior over the log($g$) ($\mathcal{U}(3,4)$). For the setup of these models, we include the radius as an additional free parameter. In Table \ref{tab:finalbest} and Figure \ref{fig:radar}, we compare the final adopted values for the M-F dataset with and without the restrictive prior. We recover the evolutionary model consistent radius and log($g$) solutions. 
In Figure \ref{fig:radar} and the corner plot of Figure \ref{fig:cornerprior}, we show that the updated $T_{\rm eff}$ is about 100\,K lower for the low log($g$) solution while the [M/H] and C/O ratio remain consistent within 3\,$\sigma$.
Our results show that it is impossible to favor any of the two model solutions statistically since their Bayesian evidence is equal within error bars (see last row in Table \ref{tab:finalbest}). Note that our restrictive prior does not degrade the error bars on other parameters, which we can interpret as high log($g$) solutions ($\sim$\,4.8\,dex) being favored from the full range because of degeneration within the model's parameters. We further explore these degeneracies in Section \ref{S.Discu}.

\section{Discussion}\label{S.Discu}

We modeled the atmosphere of AF\,Lep\,b with forward modeling and recovered the atmospheric parameters. 
We then identified that using a prior is a straightforward solution to reconcile evolutionary models with atmospheric ones. 
Our atmospheric physical properties solutions are given by the model with the restrictive prior (see Table \ref{tab:finalbest} and model b0 in Table \ref{tab:loo_b}), meaning that AF\,Lep\,b's $T_{\rm eff}$ is $\sim$\,750\,K, log($g$) is $\sim$\,3.7\,dex, [M/H] is $>$\,0.4, C/O is $\sim$\,0.6, $R$ is $\sim$\,1.2\,R$\mathrm{_{Jup}}$, and log(L/L$\mathrm{_{\odot}}$) is $\sim$\,-5.36.

The use of a prior allowed our results based on forward modeling analysis to be consistent for the $T_{\rm eff}$ and log($g$) with the previous characterization using atmospheric retrieval by \citet{Zhang2023ELementalAtmospheres}, where they derived a $T_{\rm eff}$ of 800\,K and a log($g$) of 3.7\,dex, together with an unconstrained C/O ratio and a super-solar metallicity. 
The metallicity obtained by \citet{Zhang2023ELementalAtmospheres} is higher ([Fe/H]$>$\,1) than ours ([M/H]$\sim$\,0.6). The labels used for the metallicity in both cases are different, but they both refer to the bulk metallicity. In \citet{Zhang2023ELementalAtmospheres}, the metallicity is explored with a uniform prior between -1 to 2, while in Exo-REM we can only explore values between -0.5 and 1. Both values indicate a metal-enriched atmosphere for AF\,Lep\,b, but need to be further investigated given the large uncertainties.

In our approach, the Bayesian evidence does not enable us to favor one model over another as the best solution.
In other words, the single number hiding behind the Bayesian evidence does not contain enough information to unveil where and how the models fail to perform well. 
To further address the models' robustness, we performed a cross-validation test similarly as presented by \citet{Welbanks2022OnExoplanets.}. 
The leave-one-out is a case of cross-validation tests where we remove a data point from the original set and evaluate the impact of that change on the posterior distribution. The process is repeated for each data point until every point has been used as a testing point. We based the application of this method on the in-progress book called "An Introduction to Bayesian Data Analysis for Cognitive Science" public through GitHub by Bruno Nicenboim, Daniel Schad, and Shravan Vasishth\footnote{Access the online book here: \url{https://vasishth.github.io/bayescogsci/book/}}. We computed the models for the M-F dataset and the Exo-REM grid in the cases with and without the restrictive log($g$) prior, leaving one point out iteratively. The 76 model runs with the corresponding best-fitted values and uncertainties are listed in the Appendix \ref{app.loo} in Table \ref{tab:loo_a} for the default prior and in Table \ref{tab:loo_b} for the restrictive log($g$) prior.

The analysis we can extract from those models does not lead to favor one model or the other but allows us to visualize the presence of outliers and critical points driving the results.
In Figure \ref{fig:loo0} and \ref{fig:loo1}, we present the variations for every parameter as a function of leaving one point out iteratively.
When the posterior value (color scale) is close to the value obtained when all points were included (first row in Tables \ref{tab:loo_a} and \ref{tab:loo_b}), the predictive capacity of the models in the specific wavelength is high, on the opposite side, when a point is assigned a very different value, it means that this point is driving the results. In other words, when that point is not included, the results change drastically, meaning that the point is an outlier in the data or carries valuable information on the atmospheric properties of the object. 

In the case of AF\,Lep\,b, we identified two critical conclusions from this analysis. 
The first one is that the photometric point from \citet{Franson2023AstrometricLep} at 3.8\,$\mu$m significantly impacts the temperature $T_{\rm eff}$ solution (see Figure \ref{fig:loo0}). When that point is taken out, the $T_{\rm eff}$ lowers by $\sim$\,100\,K. In the same way, this point is favoring high-metallicity solutions. In Figure \ref{fig:loo1} with a log($g$) exploration restricted, we find that the same behavior is observed. Although the new solutions for $T_{\rm eff}$, log($g$), and $R$ seem more constrained, the 3.8\,$\mu$m photometric point still drives and favors high-metallicity and super solar C/O ratio solutions. We draw attention to these results, as one single photometric point significantly influences current fitting solutions on metallicity and C/O ratios.
There can be several explanations for such a particular behavior regarding this data point.
The effect of clouds varies as we transition from near-infrared (IR) to mid-IR since the emission will reach us from different atmospheric layers \citep{Madhusudhan2019ExoplanetaryProspects}, and we cannot rule out that models fail in simultaneously handling these two regimes well. 

The second conclusion is that, when we focus on the last data point of the SPHERE/IFS spectrum from \citet{Mesa2023AFData} ($\lambda = 1.645 \mu$m), we observe that this value behaves differently with respect to its neighbors. 
The flux decreases for this last point, which does not seem realistic from the spectral energy distribution for a late-L object (see Figure \ref{fig:loo1}). This behavior can also be observed in the best model's extrapolation compared to the data points in Figure \ref{fig:FitExo}. Together, both things seem to be telling us that this point is probably an outlier that needs to be investigated with further data analysis and new observations.

\section{Conclusions}\label{S.Concl}

Our atmospheric forward-modeling analysis strongly supports the planetary nature of AF\,Lep\,b. Its spectral energy distribution between 0.96 and 3.8\,$\mu$m is consistent with that of a young, cold, early-T super-Jovian planet. 
We recover physically consistent solutions for the radius by implementing a restrictive prior on the surface gravity, which allows us to solve the small radii problem for this specific exoplanet. 
Our final solutions agree with the atmospheric retrieval solutions published by \citet{Zhang2023ELementalAtmospheres}. 
We implemented a cross-validation analysis, allowing us to identify the most critical data points driving the solutions, and we conclude that the high metallicity and super-solar carbon-to-oxygen ratio solutions driven by the photometry point from \citet{Franson2023AstrometricLep}.

Note that spectroscopic data at longer wavelengths or higher spectral resolution have the potential to solve some of the degeneracies encountered with low-resolution observations. This can be observed, for example, in the works done with VLTI/GRAVITY observations by \citet{Nowak2020AstrophysicsInterferometry} on $\beta$\,Pic\,b and by \citet{Molliere2020Retrieving8799e} on HR\,8799\,e, or with Keck/KPIC observations as published by \citet{Wang2021DetectionSpectroscopy} for HR\,8799\,cde and by \citet{Xuan2022ASpectroscopy} for HD\,4747\,B, or soon with the promising VLT/HiRISE observations as shown by \citet{Vigan2023FirstExoplanets}.
To close this work, we want to state that new observations are mandatory before concluding about AF\,Lep\,b's metal-rich nature and super-solar C/O ratio.

\begin{acknowledgements}

We would like to thank Vivien Parmentier, Andrea Chiavassa, and Mark Hammond for their insightful comments on this analysis. This publication used the SIMBAD and VizieR databases operated at the CDS, Strasbourg, France. This work has made use of data from the European Space Agency (ESA) mission Gaia \url{https://www.cosmos.esa.int/gaia}, processed by the Gaia Data Processing and Analysis Consortium (DPAC, \url{https://www.cosmos.esa.int/web/gaia/dpac/consortium}). Funding for the DPAC has been provided by national institutions, in particular, the institutions participating in the Gaia Multilateral Agreement.
P.P.-B., G.C., and M.B. received funding from the French Programme National de Planétologie (PNP) and de Physique Stellaire (PNPS) of CNRS (INSU). D.B., P.P.-B., and G.C. also received funding from the 2023 EPEx program of Observatoire Côte d'Azur. 
D.M. acknowledges the PRIN-INAF 2019 “Planetary systems at young ages (PLATEA)” and ASI-INAF agreement n.2018-16-HH.0.
C.D. acknowledges support from the European Research Council under the European Union’s Horizon 2020 research and innovation program under grant agreement No. 832428-Origins and support from Labex OSUG (Investissements d’avenir – ANR10 LABX56).
S.P. acknowledges the support of ANID, --Millennium Science Initiative Program-- Center Code NCN19\_171 and NCN2021\_080. 

\end{acknowledgements}

\bibliography{main}

\begin{thebibliography}{35}
\expandafter\ifx\csname natexlab\endcsname\relax\def\natexlab#1{#1}\fi

\bibitem[{Asplund {et~al.}(2009)Asplund, Grevesse, Sauval, \&
  Scott}]{Asplund2009TheSun}
Asplund, M., Grevesse, N., Sauval, A.~J., \& Scott, P. 2009, The Annual Review
  of Astronomy and Astrophysics, 47, 481

\bibitem[{Baudino {et~al.}(2015)Baudino, B{\'{e}}zard, Boccaletti, Bonnefoy,
  Lagrange, \& Galicher}]{baudino2015interpreting}
Baudino, J.-L., B{\'{e}}zard, B., Boccaletti, A., {et~al.} 2015, Astronomy {\&}
  Astrophysics, 582, A83

\bibitem[{Bell {et~al.}(2015)Bell, Mamajek, \& Naylor}]{Bell2015ANeighbourhood}
Bell, C. P.~M., Mamajek, E.~E., \& Naylor, T. 2015, Monthly Notices of the
  Royal Astronomical Society, 454, 593

\bibitem[{Blain {et~al.}(2021)Blain, Charnay, \& B{\'{e}}zard}]{refId0}
Blain, D., Charnay, B., \& B{\'{e}}zard, B. 2021, Astronomy {\&} Astrophysics,
  646, A15

\bibitem[{Boccaletti {et~al.}(2023)Boccaletti, M{\^{a}}lin, Baudoz, Tremplin,
  Perrot, Rouan, O., Whiteford, Molli{\`{e}}re, Waters, Henning, Decin,
  G{\"{u}}del, Vadenbussche, Absil, Argyriou, Bouwman, Cossou, Coulais,
  Gastaud, Glasse, Glauser, Kamp, Kendrew, Krause, Lahuis, Mueller, Olofsson,
  Patapis, Pye, Royer, Serabyn, Scheithauer, Colina, van Dischoeck~E., Ostlin,
  T., \& G}]{Boccaletti2023ImagingCoronagraph}
Boccaletti, A., M{\^{a}}lin, M., Baudoz, P., {et~al.} 2023, arXiv e-prints,
  arXiv:2310.13414

\bibitem[{Brandt {et~al.}(2021)Brandt, Michalik, Brandt, Li, Dupuy, \&
  Zeng}]{Brandt2021Htof:Missions}
Brandt, G.~M., Michalik, D., Brandt, T.~D., {et~al.} 2021, The Astronomical
  Journal, 162, 230

\bibitem[{Carter {et~al.}(2023)Carter, Hinkley, Kammerer, Skemer, Biller,
  Leisenring, Millar-Blanchaer, Petrus, Stone, Ward-Duong, Wang, Girard, Hines,
  Perrin, Pueyo, Balmer, Bonavita, Bonnefoy, Chauvin, Choquet, Christiaens,
  Danielski, Kennedy, Matthews, Miles, Patapis, Ray, Rickman, Sallum,
  Stapelfeldt, Whiteford, Zhou, Absil, Boccaletti, Booth, Bowler, Chen, Currie,
  Fortney, Grady, Greebaum, Henning, Hoch, Janson, Kalas, Kenworthy, Kervella,
  Kraus, Lagage, Liu, Macintosh, Marino, Marley, Marois, Matthews, Mawet,
  McElwain, Metchev, Meyer, Molliere, Moran, Morley, Mukherjee, Pantin,
  Quirrenbach, Rebollido, Ren, Schneider, Vasist, Worthen, Wyatt,
  Briesemeister, Bryan, Calissendorff, Cantalloube, Cugno, De~Furio, Dupuy,
  Factor, Faherty, Fitzgerald, Franson, Gonzales, Hood, Howe, Kuzuhara,
  Lagrange, Lawson, Lazzoni, Lew, Liu, Llop-Sayson, Lloyd, Martinez, Mazoyer,
  Palma-Bifani, Quanz, Redai, Samland, Schlieder, Tamura, Tan, Uyama, Vigan,
  Vos, Wagner, Wolff, Ygouf, Zhang, Zhang, \& Zhang}]{Carter2023Them}
Carter, A.~L., Hinkley, S., Kammerer, J., {et~al.} 2023, The Astrophysical
  Journal Letters, 951, L20

\bibitem[{Charbonneau {et~al.}(2002)Charbonneau, Brown, Noyes, \&
  Gilliland}]{Charbonneau2002DetectionAtmosphere}
Charbonneau, D., Brown, T.~M., Noyes, R.~W., \& Gilliland, R.~L. 2002, The
  Astrophysical Journal, 568, 377

\bibitem[{Charnay {et~al.}(2018)Charnay, B{\'{e}}zard, Baudino, Bonnefoy,
  Boccaletti, \& Galicher}]{charnay2018self}
Charnay, B., B{\'{e}}zard, B., Baudino, J.-L., {et~al.} 2018, The Astrophysical
  Journal, 854, 172

\bibitem[{Charnay {et~al.}(2021)Charnay, Blain, B{\'{e}}zard, Leconte, Turbet,
  \& Falco}]{Charnay2021FormationK2-18b}
Charnay, B., Blain, D., B{\'{e}}zard, B., {et~al.} 2021, Astronomy and
  Astrophysics, 646, 1

\bibitem[{Chinchilla {et~al.}(2021)Chinchilla, B{\'{e}}jar, Lodieu,
  Zapatero~Osorio, \& Gauza}]{Chinchilla2021StrongC}
Chinchilla, P., B{\'{e}}jar, V. J.~S., Lodieu, N., Zapatero~Osorio, M.~R., \&
  Gauza, B. 2021, Astronomy {\&} Astrophysics, 645, A17

\bibitem[{Currie {et~al.}(2022)Currie, Biller, Lagrange, Marois, Guyon,
  Nielsen, Bonnefoy, \& De~Rosa}]{Currie2022DirectPlanets}
Currie, T., Biller, B., Lagrange, A.-M., {et~al.} 2022, in Protostars and
  Planets, in press. edn., ed. {Shu-ichiro Inutsuka}, {Yuri Aikawa}, {Takayuki
  Muto}, {Kengo Tomida}, \& {Motohide Tamura}, Vol. VII

\bibitem[{De~Rosa {et~al.}(2023)De~Rosa, Nielsen, Wahhaj, Ruffio, Kalas, Peck,
  Hirsch, \& Roberson}]{DeRosa2023DirectLeporis}
De~Rosa, R.~J., Nielsen, E.~L., Wahhaj, Z., {et~al.} 2023, Astronomy {\&}
  Astrophysics, 672, A94

\bibitem[{Franson {et~al.}(2023)Franson, Bowler, Zhou, Pearce,
  Bardalez~Gagliuffi, Biddle, Brandt, Crepp, Dupuy, Faherty, Jensen-Clem,
  Morgan, Sanghi, Theissen, Tran, \& Wolf}]{Franson2023AstrometricLep}
Franson, K., Bowler, B.~P., Zhou, Y., {et~al.} 2023, The Astrophysical Journal
  Letters, 950, L19

\bibitem[{Galicher {et~al.}(2018)Galicher, Boccaletti, Mesa, Delorme, Gratton,
  Langlois, Lagrange, Maire, Le~Coroller, Chauvin, Biller, Cantalloube, Janson,
  Lagadec, Meunier, Vigan, Hagelberg, Bonnefoy, Zurlo, Rocha, Maurel, Jaquet,
  Buey, \& Weber}]{Galicher2018AstrometricSpeCal}
Galicher, R., Boccaletti, A., Mesa, D., {et~al.} 2018, Astronomy {\&}
  Astrophysics, 615, A92

\bibitem[{Guillot {et~al.}(2022)Guillot, Fletcher, Helled, Ikoma, Line, \&
  Parmentier}]{Guillot2022GiantInside-Out}
Guillot, T., Fletcher, L.~N., Helled, R., {et~al.} 2022, in Protostars and
  Planets, 1st edn., ed. {Shu-ichiro Inutsuka}, {Yuri Aikawa}, {Takayuki Muto},
  {Kengo Tomida}, \& {Motohide Tamura}, Vol. VII

\bibitem[{Kervella {et~al.}(2022)Kervella, Arenou, \&
  Th{\'{e}}venin}]{Kervella2022StellarEDR3}
Kervella, P., Arenou, F., \& Th{\'{e}}venin, F. 2022, Astronomy {\&}
  Astrophysics, 657, A7

\bibitem[{Lodders(2010)}]{Lodders2010ExoplanetChemistry}
Lodders, K. 2010, in Formation and Evolution of Exoplanets (Weinheim, Germany:
  Wiley-VCH Verlag GmbH {\&} Co. KGaA), 157--186

\bibitem[{Madhusudhan(2019)}]{Madhusudhan2019ExoplanetaryProspects}
Madhusudhan, N. 2019, Annual Review of Astronomy and Astrophysics, 57, 617

\bibitem[{Marleau {et~al.}(2019)Marleau, Coleman, Leleu, \&
  Mordasini}]{Marleau2019ExploringB}
Marleau, G.~D., Coleman, G.~A., Leleu, A., \& Mordasini, C. 2019, Astronomy and
  Astrophysics, 624

\bibitem[{Mesa {et~al.}(2023)Mesa, Gratton, Kervella, Bonavita, Desidera,
  D’Orazi, Marino, Zurlo, \& Rigliaco}]{Mesa2023AFData}
Mesa, D., Gratton, R., Kervella, P., {et~al.} 2023, Astronomy {\&}
  Astrophysics, 672, A93

\bibitem[{Molli{\`{e}}re {et~al.}(2020)Molli{\`{e}}re, Stolker, Lacour, Otten,
  Shangguan, Charnay, Molyarova, Nowak, Henning, Marleau, Semenov, van
  Dishoeck, Eisenhauer, Garcia, Garcia~Lopez, Girard, Greenbaum, Hinkley,
  Kervella, Kreidberg, Maire, Nasedkin, Pueyo, Snellen, Vigan, Wang, de~Zeeuw,
  \& Zurlo}]{Molliere2020Retrieving8799e}
Molli{\`{e}}re, P., Stolker, T., Lacour, S., {et~al.} 2020, Astronomy {\&}
  Astrophysics, 640, A131

\bibitem[{Nowak {et~al.}(2020)Nowak, Lacour, Molli{\`{e}}re, Wang, Charnay,
  Buron, Cantalloube, Collin, Chapron, Cl{\'{e}}net, Foresto, Zeeuw, Dembet,
  Gao, Gendron, Genzel, Gillessen, Hau{\ss}mann, Henning, Hippler, Hubert, \&
  Jocou}]{Nowak2020AstrophysicsInterferometry}
Nowak, G. C.~M., Lacour, S., Molli{\`{e}}re, P., {et~al.} 2020, 110, 1

\bibitem[{Palma-Bifani {et~al.}(2023)Palma-Bifani, Chauvin, Bonnefoy, Rojo,
  Petrus, Rodet, Langlois, Allard, Charnay, Desgrange, Homeier, Lagrange,
  Beuzit, Baudoz, Boccaletti, Chomez, Delorme, Desidera, Feldt, Ginski,
  Gratton, Maire, Meyer, Samland, Snellen, Vigan, \&
  Zhang}]{Palma-Bifani2023PeeringPic}
Palma-Bifani, P., Chauvin, G., Bonnefoy, M., {et~al.} 2023, Astronomy {\&}
  Astrophysics, 670, A90

\bibitem[{Pearce {et~al.}(2022)Pearce, Launhardt, Ostermann, Kennedy, Gennaro,
  Booth, Krivov, Cugno, Henning, Quirrenbach, Barcucci, Matthews, Ruh, \&
  Stone}]{Pearce2022PlanetDiscs}
Pearce, T.~D., Launhardt, R., Ostermann, R., {et~al.} 2022, Astronomy {\&}
  Astrophysics, 659, A135

\bibitem[{Petrus {et~al.}(2021)Petrus, Bonnefoy, Chauvin, Charnay, Marleau,
  Gratton, Lagrange, Rameau, Mordasini, Nowak, Delorme, Boccaletti, Carlotti,
  Houll{\'{e}}, Vigan, Allard, Desidera, D’Orazi, Hoeijmakers, Wyttenbach, \&
  Lavie}]{Petrus2021Medium-resolutionB}
Petrus, S., Bonnefoy, M., Chauvin, G., {et~al.} 2021, Astronomy {\&}
  Astrophysics, 648, A59

\bibitem[{Petrus {et~al.}(2023)Petrus, Chauvin, Bonnefoy, Tremblin, Charnay,
  Delorme, Marleau, Bayo, Manjavacas, Lagrange, Molli{\`{e}}re, Palma-Bifani,
  Biller, Jenkins, Goyal, \& Hoch}]{Petrus2023X-SHYNE:Analogs}
Petrus, S., Chauvin, G., Bonnefoy, M., {et~al.} 2023, Astronomy {\&}
  Astrophysics, 670, L9

\bibitem[{Pueyo(2016)}]{Pueyo2016DETECTIONMODELING}
Pueyo, L. 2016, The Astrophysical Journal, 824, 117

\bibitem[{Skilling(2006)}]{Skilling2006Skilling833-860}
Skilling, J. 2006, Bayesian Analysis, 1, 833

\bibitem[{Vigan {et~al.}(2023)Vigan, El~Morsy, Lopez, Otten, Garcia, Costes,
  Muslimov, Viret, Charles, Zins, Murray, Costille, Paufique, Seemann,
  Houll{\'{e}}, Anwand-Heerwart, Phillips, Abinanti, Balard, Baraffe,
  Benedetti, Blanchard, Blanco, Beuzit, Choquet, Cristofari, Desidera, Dohlen,
  Dorn, Ely, Fuenteseca, Garcia, Jaquet, Jaubert, Kasper, Le~Merrer, Maire,
  N'Diaye, Pallanca, Popovic, Pourcelot, Reiners, Rochat, Sehim, Schmutzer,
  Smette, Tchoubaklian, Tomlinson, \&
  Valenzuela~Soto}]{Vigan2023FirstExoplanets}
Vigan, A., El~Morsy, M., Lopez, M., {et~al.} 2023, arXiv e-prints,
  arXiv:2309.12390

\bibitem[{Wang {et~al.}(2021{\natexlab{a}})Wang, Ruffio, Morris, Delorme,
  Jovanovic, Pezzato, Echeverri, Finnerty, Hood, Zanazzi, Bryan, Bond, Cetre,
  Martin, Mawet, Skemer, Baker, Xuan, Wallace, Wang, Bartos, Blake, Boden,
  Buzard, Calvin, Chun, Doppmann, Dupuy, Duch{\^{e}}ne, Feng, Fitzgerald,
  Fortney, Freedman, Knutson, Konopacky, Lilley, Liu, Lopez, Lupu, Marley,
  Meshkat, Miles, Millar-Blanchaer, Ragland, Roy, Ruane, Sappey, Schofield,
  Weiss, Wetherell, Wizinowich, \& Ygouf}]{Wang2021DetectionSpectroscopy}
Wang, J.~J., Ruffio, J.-B., Morris, E., {et~al.} 2021{\natexlab{a}}, The
  Astronomical Journal, 162, 148

\bibitem[{Wang {et~al.}(2021{\natexlab{b}})Wang, Vigan, Lacour, Nowak, Stolker,
  De~Rosa, Ginzburg, Gao, Abuter, Amorim, Asensio-Torres, Baub{\"{o}}ck,
  Benisty, Berger, Beust, Beuzit, Blunt, Boccaletti, Bohn, Bonnefoy, Bonnet,
  Brandner, Cantalloube, Caselli, Charnay, Chauvin, Choquet, Christiaens,
  Cl{\'{e}}net, du~Foresto, Cridland, Zeeuw, Dembet, Dexter, Drescher, Duvert,
  Eckart, Eisenhauer, Facchini, Gao, Garcia, Lopez, Gardner, Gendron, Genzel,
  Gillessen, Girard, Haubois, Hei{\ss}el, Henning, Hinkley, Hippler, Horrobin,
  Houll{\'{e}}, Hubert, Jim{\'{e}}nez-Rosales, Jocou, Kammerer, Keppler,
  Kervella, Meyer, Kreidberg, Lagrange, Lapeyr{\`{e}}re, Bouquin, L{\'{e}}na,
  Lutz, Maire, M{\'{e}}nard, M{\'{e}}rand, Molli{\`{e}}re, Monnier, Mouillet,
  M{\"{u}}ller, Nasedkin, Ott, Otten, Paladini, Paumard, Perraut, Perrin,
  Pfuhl, Pueyo, Rameau, Rodet, Rodr{\'{i}}guez-Coira, Rousset, Scheithauer,
  Shangguan, Shimizu, Stadler, Straub, Straubmeier, Sturm, Tacconi, Dishoeck,
  Vincent, Fellenberg, Ward-Duong, Widmann, Wieprecht, Wiezorrek, \&
  Woillez}]{Wang2021Constrainingsup/sup}
Wang, J.~J., Vigan, A., Lacour, S., {et~al.} 2021{\natexlab{b}}, The
  Astronomical Journal, 161, 148

\bibitem[{Welbanks {et~al.}(2022)Welbanks, McGill, Line, \&
  Madhusudhan}]{Welbanks2022OnExoplanets.}
Welbanks, L., McGill, P., Line, M., \& Madhusudhan, N. 2022, in Bulletin of the
  American Astronomical Society, Vol.~54, 102.105

\bibitem[{Xuan {et~al.}(2022)Xuan, Wang, Ruffio, Knutson, Mawet,
  Molli{\`{e}}re, Kolecki, Vigan, Mukherjee, Wallack, Wang, Baker, Bartos,
  Blake, Bond, Bryan, Calvin, Cetre, Chun, Delorme, Doppmann, Echeverri,
  Finnerty, Fitzgerald, Horstman, Inglis, Jovanovic, L{\'{o}}pez, Martin,
  Morris, Pezzato, Ragland, Ren, Ruane, Sappey, Schofield, Skemer, Venenciano,
  Wallace, \& Wizinowich}]{Xuan2022ASpectroscopy}
Xuan, J.~W., Wang, J., Ruffio, J.-B., {et~al.} 2022, The Astrophysical Journal,
  937, 54

\bibitem[{Zhang {et~al.}(2023)Zhang, Molli{\`{e}}re, Hawkins, Manea, Fortney,
  Morley, Skemer, Marley, Bowler, Carter, Franson, Maas, \&
  Sneden}]{Zhang2023ELementalAtmospheres}
Zhang, Z., Molli{\`{e}}re, P., Hawkins, K., {et~al.} 2023, The Astronomical
  Journal, 166, 198

\end{thebibliography}
\bibliographystyle{aa}

\begin{appendix}
\onecolumn
\section{Details on the atmospheric modeling with \texttt{ForMoSA}}\label{app.models}

The posterior best values and 1\,$\sigma$ asymmetric confidence intervals for each free parameter and dataset can be found in Table \ref{tab:ForMoSA}. 
We also list the Bayesian evidence used to compare models in the last column of Table \ref{tab:ForMoSA}. 

\renewcommand{\arraystretch}{1.8}
\begin{table*}[h]
\caption{Posteriors of the atmospheric models performed with \texttt{ForMoSA}.}
\centering
\resizebox{\textwidth}{!}{
\begin{tabular}{|l : c c c c c c c : c c |}
\hline
Label & $T_{\rm eff}$ (K) & log($g$) (dex) & [M/H] & C/O & R (R$\mathrm{_{Jup}}$) & log(L/L$\mathrm{_{\odot}}$) & Av (mag) & log(z) & h \\  
\hline\hline
Priors(E)&$\mathcal{U}(400,2000)$&$\mathcal{U}(3,5)$&$\mathcal{U}(-0.5,1)$&$\mathcal{U}(0.1,0.8)$&$\mathcal{U}(0.1,10)$& - &$\mathcal{U}(0,20)$&&\\
\hline
\multicolumn{9}{l}{Dataset DR-F: \citet{DeRosa2023DirectLeporis} and \citet{Franson2023AstrometricLep}}\\
\hline
\teal{DR-F E1}&$902^{+16}_{-16}$&$4.4^{+0.1}_{-0.1}$&$0.46^{+0.04}_{-0.06}$&$0.50^{+0.01}_{-0.01}$&-&-&-&$-20.2\pm0.1$&$8.4$\\ 
DR-F E2&$902^{+15}_{-13}$&$4.4^{+0.1}_{-0.1}$&$0.44^{+0.04}_{-0.06}$&$0.50^{+0.01}_{-0.01}$&$0.9^{+0.1}_{-0.1}$&$-5.35^{+0.02}_{-0.03}$&-&$-25.6\pm0.2$&$13.4$\\ 
DR-F E3&$1871^{+92}_{-116}$&$3.9^{+0.4}_{-0.4}$&$-0.50^{+0.08}_{-0.04}$&$0.71^{+0.1}_{-0.23}$&$0.3^{+0.1}_{-0.1}$&$-4.88^{+0.06}_{-0.07}$&$7.18^{+1.01}_{-1.26}$&$-26.0\pm0.2$&$11.5$\\ 
\hline
\multicolumn{9}{l}{Dataset M-F: \citet{Mesa2023AFData} and \citet{Franson2023AstrometricLep}}\\
\hline

\teal{M-F E1}&$836^{+71}_{-86}$&$4.9^{+0.1}_{-0.7}$&$0.67^{+0.08}_{-0.13}$&$0.51^{+0.1}_{-0.14}$&-&-&-&$-10.8\pm0.1$&$4.3$\\ 
M-F E2&$833^{+69}_{-99}$&$4.9^{+0.1}_{-1.0}$&$0.68^{+0.07}_{-0.14}$&$0.51^{+0.12}_{-0.12}$&$0.9^{+0.2}_{-0.2}$&$-5.40^{+0.09}_{-0.07}$&-&$-16.0\pm0.2$&$8.9$\\ 
M-F E3&$902^{+78}_{-85}$&$4.6^{+0.2}_{-0.5}$&$0.65^{+0.09}_{-0.22}$&$0.52^{+0.11}_{-0.14}$&$0.8^{+0.2}_{-0.1}$&$-5.38^{+0.06}_{-0.06}$&$0.51^{+0.92}_{-0.49}$&$-18.8\pm0.2$&$10.0$\\ 
\hline
\end{tabular}
}
\tablefoot{Priors, posteriors, and adopted values for the models described in Section \ref{S.Atmo}. The table is divided into priors, DR-F dataset posteriors, and M-F dataset posteriors. The labels' first character refers to the atmospheric model (A for ATMO and E for Exo-REM), while the second corresponds to the parameter setup, from 1 to 5. The best model label (highest log(z) values) are colored.}
\label{tab:ForMoSA}
\end{table*}

\section{Leave-one-out posteriors}\label{app.loo}

The two Tables in this section show the same results as Table \ref{tab:ForMoSA} but for the leave-one-out setup. The M-F dataset has 37 points in total. We, therefore, made 38 models in both cases, where one point was left out iteratively, and the model with index 0 is the validation one with all points included.

\begin{figure*}[ht!]
\centering
\includegraphics[width=\hsize]{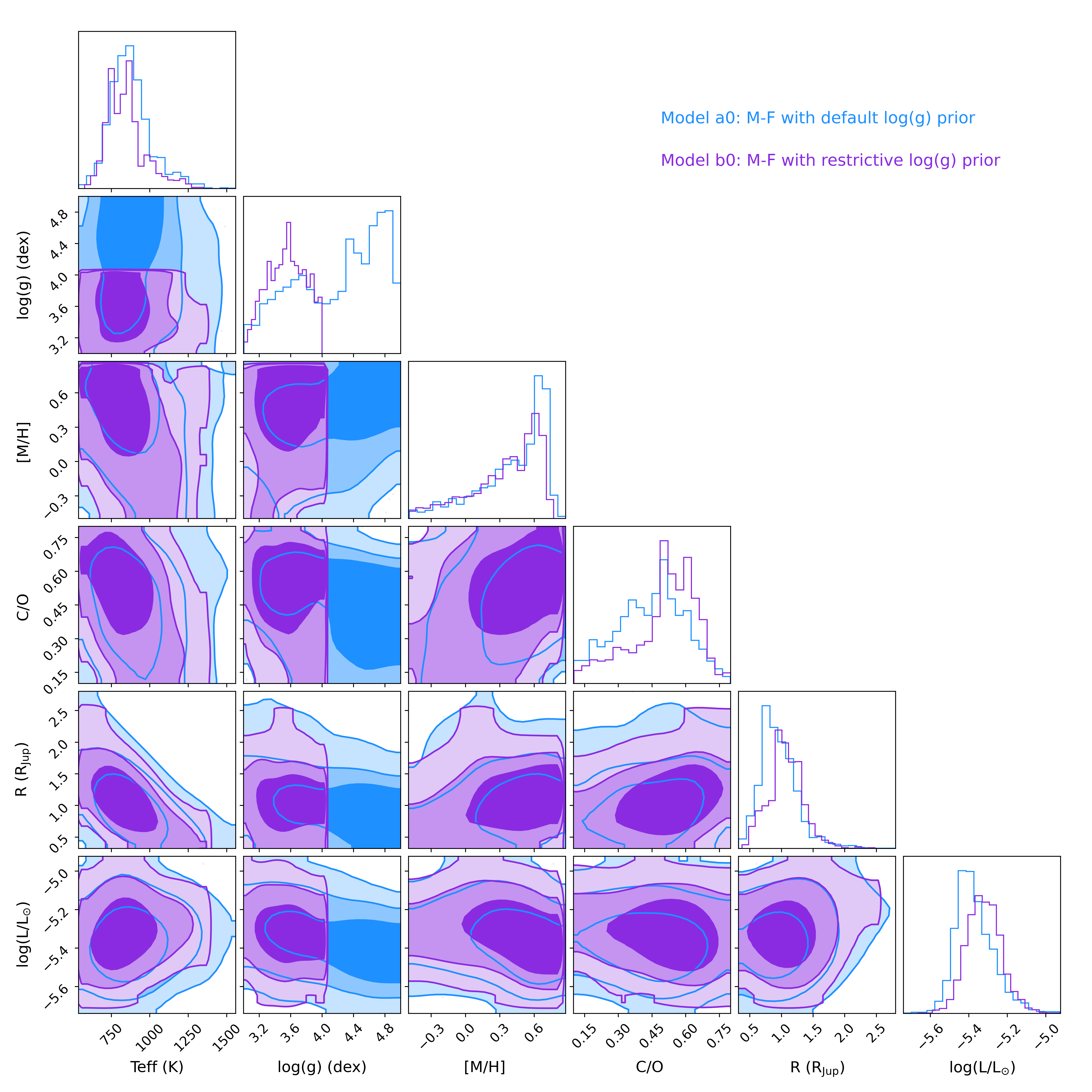}
    \caption{Cornet plot comparing the posteriors distributions for the models a0 and b0 from Table \ref{tab:loo_a} and \ref{tab:loo_b} where a $\mathcal{U}(3,4)$ prior was added to the purple model (b0) to recover solutions in agreement with the evolutionary models.}
\label{fig:cornerprior}
\end{figure*}

\begin{table*}[ht]
\caption{Leave-one-out models for the M-F dataset with Exo-REM and default priors.}
\centering
\resizebox{0.7\textwidth}{!}{
\begin{tabular}{|l : c c c c c c : c c |}
\hline
Label & $T_{\rm eff}$ (K) & log($g$) (dex) & [M/H] & C/O & R (R$\mathrm{_{Jup}}$) & log(L/L$\mathrm{_{\odot}}$) & log(z) & h \\  
\hline\hline
Priors&$\mathcal{U}(400,2000)$&$\mathcal{U}(3,5)$&$\mathcal{U}(-0.5,1)$&$\mathcal{U}(0.1,0.8)$&$\mathcal{U}(0.1,10)$&&&\\
\hline\hline
a0&$854.0^{+63.0}_{-81.0}$&$4.97^{+0.17}_{-0.34}$&$0.69^{+0.04}_{-0.06}$&$0.49^{+0.08}_{-0.12}$&$0.84^{+0.23}_{-0.11}$&$-5.43^{+0.05}_{-0.04}$&$-16.7\pm0.2$&$9.5$\\ 
\hline\hline
a1&$835.0^{+59.0}_{-65.0}$&$4.94^{+0.2}_{-0.35}$&$0.69^{+0.05}_{-0.06}$&$0.51^{+0.09}_{-0.13}$&$0.86^{+0.21}_{-0.11}$&$-5.43^{+0.06}_{-0.04}$&$-16.4\pm0.2$&$9.2$\\ 
a2&$826.0^{+65.0}_{-77.0}$&$4.99^{+0.18}_{-0.36}$&$0.69^{+0.05}_{-0.08}$&$0.51^{+0.08}_{-0.12}$&$0.87^{+0.22}_{-0.11}$&$-5.43^{+0.05}_{-0.05}$&$-16.1\pm0.2$&$8.8$\\ 
a3&$836.0^{+66.0}_{-79.0}$&$4.97^{+0.2}_{-0.47}$&$0.7^{+0.05}_{-0.06}$&$0.5^{+0.09}_{-0.12}$&$0.86^{+0.22}_{-0.14}$&$-5.44^{+0.06}_{-0.05}$&$-16.2\pm0.2$&$8.8$\\ 
a4&$860.0^{+57.0}_{-66.0}$&$4.98^{+0.18}_{-0.35}$&$0.7^{+0.05}_{-0.05}$&$0.49^{+0.08}_{-0.12}$&$0.82^{+0.19}_{-0.12}$&$-5.44^{+0.06}_{-0.04}$&$-15.9\pm0.2$&$8.9$\\ 
a5&$871.0^{+63.0}_{-74.0}$&$4.87^{+0.22}_{-0.36}$&$0.69^{+0.05}_{-0.05}$&$0.5^{+0.09}_{-0.12}$&$0.8^{+0.24}_{-0.12}$&$-5.44^{+0.06}_{-0.04}$&$-16.1\pm0.2$&$9.0$\\ 
a6&$889.0^{+59.0}_{-83.0}$&$4.76^{+0.18}_{-0.45}$&$0.69^{+0.06}_{-0.06}$&$0.51^{+0.09}_{-0.11}$&$0.77^{+0.21}_{-0.12}$&$-5.43^{+0.06}_{-0.05}$&$-15.7\pm0.2$&$8.4$\\ 
a7&$826.0^{+66.0}_{-70.0}$&$4.98^{+0.21}_{-0.4}$&$0.7^{+0.05}_{-0.06}$&$0.5^{+0.1}_{-0.11}$&$0.87^{+0.2}_{-0.12}$&$-5.44^{+0.06}_{-0.05}$&$-16.1\pm0.2$&$8.9$\\ 
a8&$817.0^{+61.0}_{-72.0}$&$4.84^{+0.2}_{-0.35}$&$0.68^{+0.05}_{-0.07}$&$0.5^{+0.08}_{-0.11}$&$0.9^{+0.22}_{-0.11}$&$-5.43^{+0.06}_{-0.05}$&$-15.8\pm0.2$&$8.6$\\ 
a9&$847.0^{+61.0}_{-80.0}$&$4.84^{+0.17}_{-0.35}$&$0.68^{+0.05}_{-0.06}$&$0.48^{+0.09}_{-0.11}$&$0.84^{+0.22}_{-0.1}$&$-5.44^{+0.06}_{-0.04}$&$-16.5\pm0.2$&$9.3$\\ 
a10&$852.0^{+58.0}_{-81.0}$&$4.97^{+0.17}_{-0.32}$&$0.69^{+0.04}_{-0.06}$&$0.49^{+0.08}_{-0.12}$&$0.84^{+0.22}_{-0.12}$&$-5.44^{+0.07}_{-0.04}$&$-16.0\pm0.2$&$8.9$\\ 
a11&$845.0^{+59.0}_{-65.0}$&$4.88^{+0.16}_{-0.33}$&$0.7^{+0.05}_{-0.07}$&$0.49^{+0.07}_{-0.12}$&$0.85^{+0.18}_{-0.11}$&$-5.44^{+0.06}_{-0.05}$&$-15.9\pm0.2$&$8.7$\\ 
a12&$848.0^{+55.0}_{-81.0}$&$4.97^{+0.21}_{-0.36}$&$0.68^{+0.05}_{-0.06}$&$0.49^{+0.09}_{-0.12}$&$0.84^{+0.24}_{-0.11}$&$-5.44^{+0.07}_{-0.04}$&$-16.0\pm0.2$&$8.7$\\ 
a13&$844.0^{+59.0}_{-77.0}$&$4.77^{+0.18}_{-0.4}$&$0.66^{+0.05}_{-0.06}$&$0.5^{+0.08}_{-0.11}$&$0.85^{+0.19}_{-0.11}$&$-5.44^{+0.06}_{-0.05}$&$-15.5\pm0.2$&$8.8$\\ 
a14&$874.0^{+62.0}_{-80.0}$&$4.93^{+0.19}_{-0.41}$&$0.69^{+0.05}_{-0.06}$&$0.49^{+0.06}_{-0.12}$&$0.8^{+0.2}_{-0.11}$&$-5.44^{+0.06}_{-0.04}$&$-16.1\pm0.2$&$8.8$\\ 
a15&$837.0^{+75.0}_{-67.0}$&$4.88^{+0.17}_{-0.31}$&$0.69^{+0.06}_{-0.05}$&$0.49^{+0.09}_{-0.12}$&$0.85^{+0.22}_{-0.12}$&$-5.44^{+0.06}_{-0.05}$&$-16.0\pm0.2$&$8.8$\\ 
a16&$889.0^{+69.0}_{-72.0}$&$4.69^{+0.21}_{-0.32}$&$0.68^{+0.05}_{-0.06}$&$0.51^{+0.1}_{-0.11}$&$0.78^{+0.19}_{-0.16}$&$-5.44^{+0.05}_{-0.05}$&$-16.0\pm0.2$&$8.9$\\ 
a17&$836.0^{+63.0}_{-87.0}$&$4.88^{+0.19}_{-0.33}$&$0.68^{+0.05}_{-0.06}$&$0.51^{+0.08}_{-0.12}$&$0.88^{+0.23}_{-0.1}$&$-5.43^{+0.05}_{-0.04}$&$-15.9\pm0.2$&$8.9$\\ 
a18&$840.0^{+59.0}_{-67.0}$&$4.9^{+0.2}_{-0.36}$&$0.7^{+0.06}_{-0.09}$&$0.5^{+0.09}_{-0.12}$&$0.87^{+0.16}_{-0.09}$&$-5.43^{+0.06}_{-0.05}$&$-15.3\pm0.2$&$8.7$\\ 
a19&$819.0^{+70.0}_{-65.0}$&$4.9^{+0.18}_{-0.36}$&$0.68^{+0.05}_{-0.07}$&$0.5^{+0.07}_{-0.12}$&$0.9^{+0.21}_{-0.12}$&$-5.44^{+0.06}_{-0.05}$&$-16.2\pm0.2$&$8.9$\\ 
a20&$866.0^{+63.0}_{-84.0}$&$4.68^{+0.19}_{-0.39}$&$0.68^{+0.05}_{-0.08}$&$0.49^{+0.1}_{-0.13}$&$0.82^{+0.22}_{-0.11}$&$-5.42^{+0.06}_{-0.05}$&$-15.6\pm0.2$&$8.5$\\ 
a21&$884.0^{+57.0}_{-68.0}$&$4.56^{+0.16}_{-0.35}$&$0.68^{+0.05}_{-0.06}$&$0.51^{+0.09}_{-0.11}$&$0.79^{+0.21}_{-0.11}$&$-5.44^{+0.06}_{-0.05}$&$-16.0\pm0.2$&$8.8$\\ 
a22&$879.0^{+54.0}_{-79.0}$&$4.77^{+0.17}_{-0.32}$&$0.69^{+0.05}_{-0.05}$&$0.5^{+0.08}_{-0.11}$&$0.8^{+0.21}_{-0.1}$&$-5.44^{+0.06}_{-0.05}$&$-16.1\pm0.2$&$9.0$\\ 
a23&$870.0^{+67.0}_{-74.0}$&$4.97^{+0.18}_{-0.34}$&$0.69^{+0.05}_{-0.05}$&$0.5^{+0.07}_{-0.13}$&$0.81^{+0.22}_{-0.12}$&$-5.44^{+0.07}_{-0.05}$&$-16.1\pm0.2$&$9.1$\\ 
a24&$859.0^{+65.0}_{-84.0}$&$4.93^{+0.25}_{-0.49}$&$0.7^{+0.05}_{-0.08}$&$0.49^{+0.1}_{-0.11}$&$0.83^{+0.2}_{-0.14}$&$-5.43^{+0.07}_{-0.05}$&$-15.9\pm0.2$&$8.9$\\ 
a25&$870.0^{+61.0}_{-60.0}$&$4.96^{+0.21}_{-0.33}$&$0.69^{+0.05}_{-0.05}$&$0.38^{+0.07}_{-0.12}$&$0.83^{+0.19}_{-0.13}$&$-5.44^{+0.07}_{-0.05}$&$-16.0\pm0.2$&$9.0$\\ 
a26&$916.0^{+76.0}_{-78.0}$&$4.58^{+0.21}_{-0.37}$&$0.69^{+0.05}_{-0.07}$&$0.51^{+0.1}_{-0.13}$&$0.74^{+0.23}_{-0.12}$&$-5.43^{+0.05}_{-0.05}$&$-16.1\pm0.2$&$8.7$\\ 
a27&$826.0^{+59.0}_{-66.0}$&$4.81^{+0.21}_{-0.35}$&$0.69^{+0.05}_{-0.06}$&$0.49^{+0.08}_{-0.12}$&$0.9^{+0.2}_{-0.13}$&$-5.43^{+0.06}_{-0.05}$&$-16.1\pm0.2$&$9.1$\\ 
a28&$868.0^{+61.0}_{-85.0}$&$4.9^{+0.17}_{-0.32}$&$0.67^{+0.05}_{-0.08}$&$0.5^{+0.08}_{-0.12}$&$0.81^{+0.21}_{-0.1}$&$-5.44^{+0.06}_{-0.05}$&$-15.8\pm0.2$&$8.8$\\ 
a29&$909.0^{+58.0}_{-65.0}$&$4.74^{+0.19}_{-0.35}$&$0.71^{+0.05}_{-0.07}$&$0.5^{+0.06}_{-0.12}$&$0.75^{+0.2}_{-0.12}$&$-5.43^{+0.07}_{-0.05}$&$-16.1\pm0.2$&$8.9$\\ 
a30&$868.0^{+69.0}_{-73.0}$&$4.95^{+0.19}_{-0.33}$&$0.69^{+0.05}_{-0.06}$&$0.49^{+0.07}_{-0.12}$&$0.8^{+0.21}_{-0.13}$&$-5.43^{+0.06}_{-0.05}$&$-16.1\pm0.2$&$8.8$\\ 
a31&$826.0^{+59.0}_{-70.0}$&$4.76^{+0.17}_{-0.39}$&$0.66^{+0.05}_{-0.06}$&$0.5^{+0.07}_{-0.12}$&$0.89^{+0.2}_{-0.11}$&$-5.43^{+0.06}_{-0.05}$&$-16.0\pm0.2$&$8.7$\\ 
a32&$801.0^{+64.0}_{-68.0}$&$4.46^{+0.2}_{-0.38}$&$0.67^{+0.05}_{-0.07}$&$0.37^{+0.09}_{-0.12}$&$1.14^{+0.2}_{-0.13}$&$-5.42^{+0.06}_{-0.06}$&$-16.0\pm0.2$&$8.8$\\ 
a33&$830.0^{+65.0}_{-73.0}$&$4.96^{+0.21}_{-0.29}$&$0.7^{+0.05}_{-0.06}$&$0.5^{+0.08}_{-0.11}$&$0.86^{+0.2}_{-0.11}$&$-5.44^{+0.06}_{-0.05}$&$-16.6\pm0.2$&$9.5$\\ 
a34&$877.0^{+46.0}_{-52.0}$&$4.62^{+0.32}_{-0.98}$&$0.68^{+0.06}_{-0.1}$&$0.51^{+0.05}_{-0.11}$&$0.81^{+0.17}_{-0.12}$&$-5.42^{+0.08}_{-0.06}$&$-15.4\pm0.2$&$8.8$\\ 
a35&$865.0^{+48.0}_{-76.0}$&$4.66^{+0.24}_{-0.57}$&$0.67^{+0.1}_{-0.26}$&$0.5^{+0.09}_{-0.18}$&$0.82^{+0.16}_{-0.15}$&$-5.41^{+0.07}_{-0.07}$&$-14.8\pm0.2$&$7.8$\\ 
a36&$744.0^{+107.0}_{-73.0}$&$4.55^{+0.43}_{-0.71}$&$0.73^{+0.04}_{-0.05}$&$0.6^{+0.08}_{-0.17}$&$1.09^{+0.23}_{-0.27}$&$-5.46^{+0.05}_{-0.04}$&$-15.1\pm0.2$&$8.7$\\ 
a37&$715.0^{+92.0}_{-72.0}$&$4.04^{+0.42}_{-0.68}$&$0.25^{+0.07}_{-0.61}$&$0.45^{+0.11}_{-0.1}$&$1.94^{+1.22}_{-0.32}$&$-5.32^{+0.37}_{-0.12}$&$-14.7\pm0.2$&$7.9$\\ 
\hline
\end{tabular}
}
\tablefoot{Priors and posteriors for the leave-one-out analysis. For model a0, all data points were included; for the rest, the point corresponding to the index number was kept out.}
\label{tab:loo_a}
\end{table*}

\begin{table*}[ht]
\caption{Leave-one-out models for the M-F dataset with Exo-REM and a restrictive log($g$) prior.}
\centering
\resizebox{0.7\textwidth}{!}{
\begin{tabular}{|l : c c c c c c : c c |}
\hline
Label & $T_{\rm eff}$ (K) & log($g$) (dex) & [M/H] & C/O & R (R$\mathrm{_{Jup}}$) & log(L/L$\mathrm{_{\odot}}$) & log(z) & h \\  
\hline\hline
Priors&$\mathcal{U}(400,2000)$&$\mathcal{U}(3,4)$&$\mathcal{U}(-0.5,1)$&$\mathcal{U}(0.1,0.8)$&$\mathcal{U}(0.1,10)$&&&\\
\hline\hline
b0&$750.0^{+99.0}_{-36.0}$&$3.68^{+0.07}_{-0.06}$&$0.6^{+0.08}_{-0.13}$&$0.61^{+0.05}_{-0.09}$&$1.23^{+0.13}_{-0.18}$&$-5.36^{+0.05}_{-0.05}$&$-16.4\pm0.2$&$8.3$\\ 
\hline\hline
b1&$754.0^{+90.0}_{-45.0}$&$3.8^{+0.22}_{-0.2}$&$0.66^{+0.07}_{-0.11}$&$0.61^{+0.05}_{-0.08}$&$1.21^{+0.19}_{-0.15}$&$-5.37^{+0.06}_{-0.05}$&$-16.8\pm0.2$&$8.7$\\ 
b2&$751.0^{+52.0}_{-90.0}$&$3.77^{+0.19}_{-0.25}$&$0.62^{+0.08}_{-0.14}$&$0.6^{+0.06}_{-0.07}$&$1.18^{+0.17}_{-0.16}$&$-5.36^{+0.05}_{-0.06}$&$-16.6\pm0.2$&$8.3$\\ 
b3&$761.0^{+43.0}_{-100.0}$&$3.86^{+0.21}_{-0.25}$&$0.66^{+0.08}_{-0.18}$&$0.6^{+0.06}_{-0.08}$&$1.13^{+0.18}_{-0.14}$&$-5.37^{+0.07}_{-0.05}$&$-16.9\pm0.2$&$8.5$\\ 
b4&$751.0^{+48.0}_{-95.0}$&$3.74^{+0.22}_{-0.28}$&$0.63^{+0.08}_{-0.15}$&$0.6^{+0.07}_{-0.07}$&$1.2^{+0.21}_{-0.14}$&$-5.36^{+0.06}_{-0.06}$&$-16.6\pm0.2$&$8.3$\\ 
b5&$743.0^{+41.0}_{-106.0}$&$3.8^{+0.21}_{-0.21}$&$0.65^{+0.09}_{-0.14}$&$0.61^{+0.08}_{-0.06}$&$1.24^{+0.22}_{-0.12}$&$-5.35^{+0.05}_{-0.06}$&$-17.0\pm0.2$&$8.6$\\ 
b6&$750.0^{+104.0}_{-35.0}$&$3.82^{+0.23}_{-0.27}$&$0.64^{+0.06}_{-0.12}$&$0.61^{+0.05}_{-0.08}$&$1.19^{+0.14}_{-0.17}$&$-5.37^{+0.06}_{-0.05}$&$-17.1\pm0.2$&$9.0$\\ 
b7&$748.0^{+91.0}_{-54.0}$&$3.86^{+0.21}_{-0.33}$&$0.69^{+0.07}_{-0.16}$&$0.6^{+0.06}_{-0.08}$&$1.2^{+0.17}_{-0.15}$&$-5.36^{+0.06}_{-0.06}$&$-16.9\pm0.2$&$8.8$\\ 
b8&$753.0^{+85.0}_{-54.0}$&$3.74^{+0.21}_{-0.27}$&$0.63^{+0.07}_{-0.17}$&$0.59^{+0.07}_{-0.08}$&$1.22^{+0.17}_{-0.15}$&$-5.36^{+0.06}_{-0.06}$&$-16.5\pm0.2$&$8.3$\\ 
b9&$754.0^{+57.0}_{-90.0}$&$3.98^{+0.23}_{-0.24}$&$0.64^{+0.07}_{-0.15}$&$0.62^{+0.06}_{-0.08}$&$1.13^{+0.2}_{-0.12}$&$-5.37^{+0.06}_{-0.06}$&$-16.7\pm0.2$&$8.6$\\ 
b10&$752.0^{+58.0}_{-85.0}$&$3.94^{+0.22}_{-0.25}$&$0.67^{+0.07}_{-0.16}$&$0.61^{+0.07}_{-0.07}$&$1.15^{+0.17}_{-0.15}$&$-5.37^{+0.06}_{-0.06}$&$-16.6\pm0.2$&$8.3$\\ 
b11&$750.0^{+90.0}_{-47.0}$&$3.8^{+0.21}_{-0.25}$&$0.65^{+0.07}_{-0.12}$&$0.6^{+0.06}_{-0.07}$&$1.21^{+0.18}_{-0.14}$&$-5.37^{+0.05}_{-0.05}$&$-17.2\pm0.2$&$8.8$\\ 
b12&$762.0^{+58.0}_{-81.0}$&$3.98^{+0.25}_{-0.3}$&$0.68^{+0.08}_{-0.15}$&$0.61^{+0.07}_{-0.08}$&$1.09^{+0.18}_{-0.15}$&$-5.36^{+0.07}_{-0.06}$&$-16.6\pm0.2$&$8.3$\\ 
b13&$747.0^{+62.0}_{-85.0}$&$3.99^{+0.21}_{-0.23}$&$0.67^{+0.08}_{-0.12}$&$0.62^{+0.05}_{-0.08}$&$1.18^{+0.18}_{-0.15}$&$-5.37^{+0.06}_{-0.06}$&$-16.1\pm0.2$&$8.3$\\ 
b14&$742.0^{+92.0}_{-51.0}$&$3.96^{+0.22}_{-0.23}$&$0.66^{+0.07}_{-0.14}$&$0.63^{+0.05}_{-0.08}$&$1.18^{+0.19}_{-0.16}$&$-5.36^{+0.06}_{-0.06}$&$-16.6\pm0.2$&$8.4$\\ 
b15&$752.0^{+50.0}_{-95.0}$&$3.76^{+0.22}_{-0.26}$&$0.63^{+0.08}_{-0.16}$&$0.59^{+0.07}_{-0.07}$&$1.22^{+0.2}_{-0.12}$&$-5.36^{+0.06}_{-0.05}$&$-16.8\pm0.2$&$8.3$\\ 
b16&$733.0^{+116.0}_{-34.0}$&$3.69^{+0.26}_{-0.26}$&$0.66^{+0.07}_{-0.13}$&$0.63^{+0.05}_{-0.09}$&$1.24^{+0.17}_{-0.2}$&$-5.37^{+0.06}_{-0.06}$&$-16.5\pm0.2$&$8.3$\\ 
b17&$745.0^{+53.0}_{-95.0}$&$3.69^{+0.19}_{-0.21}$&$0.61^{+0.08}_{-0.18}$&$0.6^{+0.07}_{-0.06}$&$1.27^{+0.18}_{-0.14}$&$-5.36^{+0.06}_{-0.06}$&$-16.7\pm0.2$&$8.3$\\ 
b18&$747.0^{+38.0}_{-104.0}$&$3.9^{+0.22}_{-0.24}$&$0.64^{+0.11}_{-0.19}$&$0.63^{+0.09}_{-0.05}$&$1.16^{+0.18}_{-0.12}$&$-5.33^{+0.06}_{-0.07}$&$-16.0\pm0.2$&$8.5$\\ 
b19&$748.0^{+105.0}_{-44.0}$&$3.77^{+0.21}_{-0.27}$&$0.64^{+0.08}_{-0.16}$&$0.6^{+0.06}_{-0.08}$&$1.25^{+0.17}_{-0.17}$&$-5.37^{+0.07}_{-0.06}$&$-16.6\pm0.2$&$8.1$\\ 
b20&$761.0^{+44.0}_{-98.0}$&$3.99^{+0.2}_{-0.24}$&$0.67^{+0.09}_{-0.16}$&$0.62^{+0.08}_{-0.06}$&$1.09^{+0.2}_{-0.11}$&$-5.35^{+0.07}_{-0.06}$&$-16.3\pm0.2$&$8.4$\\ 
b21&$757.0^{+97.0}_{-45.0}$&$3.73^{+0.23}_{-0.27}$&$0.63^{+0.07}_{-0.15}$&$0.6^{+0.07}_{-0.08}$&$1.21^{+0.16}_{-0.17}$&$-5.36^{+0.06}_{-0.06}$&$-16.8\pm0.2$&$8.5$\\ 
b22&$751.0^{+52.0}_{-94.0}$&$3.7^{+0.23}_{-0.26}$&$0.6^{+0.08}_{-0.18}$&$0.61^{+0.06}_{-0.07}$&$1.22^{+0.18}_{-0.15}$&$-5.36^{+0.07}_{-0.07}$&$-17.0\pm0.2$&$8.5$\\ 
b23&$755.0^{+43.0}_{-103.0}$&$3.77^{+0.23}_{-0.24}$&$0.64^{+0.07}_{-0.13}$&$0.6^{+0.05}_{-0.09}$&$1.21^{+0.17}_{-0.15}$&$-5.37^{+0.06}_{-0.05}$&$-16.5\pm0.2$&$8.4$\\ 
b24&$752.0^{+43.0}_{-106.0}$&$3.92^{+0.25}_{-0.22}$&$0.64^{+0.09}_{-0.18}$&$0.62^{+0.08}_{-0.07}$&$1.17^{+0.19}_{-0.14}$&$-5.35^{+0.06}_{-0.07}$&$-16.5\pm0.2$&$8.3$\\ 
b25&$758.0^{+92.0}_{-53.0}$&$3.94^{+0.26}_{-0.27}$&$0.66^{+0.08}_{-0.14}$&$0.62^{+0.07}_{-0.08}$&$1.15^{+0.17}_{-0.16}$&$-5.36^{+0.06}_{-0.06}$&$-16.6\pm0.2$&$8.3$\\ 
b26&$748.0^{+96.0}_{-42.0}$&$3.69^{+0.21}_{-0.27}$&$0.62^{+0.07}_{-0.15}$&$0.6^{+0.05}_{-0.08}$&$1.25^{+0.17}_{-0.14}$&$-5.36^{+0.06}_{-0.06}$&$-16.5\pm0.2$&$8.2$\\ 
b27&$749.0^{+45.0}_{-95.0}$&$3.84^{+0.23}_{-0.25}$&$0.66^{+0.07}_{-0.12}$&$0.61^{+0.06}_{-0.08}$&$1.18^{+0.19}_{-0.15}$&$-5.37^{+0.06}_{-0.06}$&$-17.0\pm0.2$&$8.6$\\ 
b28&$756.0^{+105.0}_{-45.0}$&$3.91^{+0.24}_{-0.25}$&$0.63^{+0.08}_{-0.18}$&$0.63^{+0.06}_{-0.09}$&$1.15^{+0.18}_{-0.17}$&$-5.36^{+0.07}_{-0.06}$&$-16.3\pm0.2$&$8.3$\\ 
b29&$762.0^{+38.0}_{-101.0}$&$3.92^{+0.22}_{-0.22}$&$0.66^{+0.08}_{-0.18}$&$0.59^{+0.1}_{-0.04}$&$1.1^{+0.19}_{-0.12}$&$-5.36^{+0.06}_{-0.06}$&$-16.7\pm0.2$&$8.4$\\ 
b30&$742.0^{+103.0}_{-46.0}$&$3.75^{+0.21}_{-0.26}$&$0.64^{+0.07}_{-0.11}$&$0.6^{+0.06}_{-0.08}$&$1.27^{+0.17}_{-0.18}$&$-5.37^{+0.06}_{-0.05}$&$-17.0\pm0.2$&$8.7$\\ 
b31&$753.0^{+91.0}_{-46.0}$&$3.71^{+0.24}_{-0.24}$&$0.62^{+0.07}_{-0.15}$&$0.6^{+0.06}_{-0.08}$&$1.22^{+0.14}_{-0.16}$&$-5.37^{+0.06}_{-0.06}$&$-16.7\pm0.2$&$8.4$\\ 
b32&$757.0^{+45.0}_{-100.0}$&$3.76^{+0.2}_{-0.26}$&$0.62^{+0.09}_{-0.18}$&$0.6^{+0.09}_{-0.06}$&$1.17^{+0.19}_{-0.13}$&$-5.36^{+0.06}_{-0.07}$&$-16.4\pm0.2$&$8.3$\\ 
b33&$751.0^{+102.0}_{-43.0}$&$3.74^{+0.22}_{-0.22}$&$0.64^{+0.08}_{-0.14}$&$0.6^{+0.05}_{-0.08}$&$1.23^{+0.15}_{-0.17}$&$-5.37^{+0.07}_{-0.06}$&$-16.6\pm0.2$&$8.3$\\ 
b34&$857.0^{+31.0}_{-92.0}$&$3.44^{+0.21}_{-0.18}$&$0.59^{+0.13}_{-0.15}$&$0.49^{+0.09}_{-0.03}$&$0.91^{+0.14}_{-0.11}$&$-5.33^{+0.06}_{-0.06}$&$-15.5\pm0.2$&$8.6$\\ 
b35&$740.0^{+77.0}_{-55.0}$&$3.77^{+0.23}_{-0.28}$&$0.65^{+0.22}_{-0.2}$&$0.6^{+0.09}_{-0.11}$&$1.23^{+0.2}_{-0.17}$&$-5.33^{+0.06}_{-0.07}$&$-15.3\pm0.2$&$7.5$\\ 
b36&$724.0^{+44.0}_{-43.0}$&$3.71^{+0.22}_{-0.36}$&$0.69^{+0.04}_{-0.04}$&$0.66^{+0.03}_{-0.04}$&$1.21^{+0.24}_{-0.15}$&$-5.46^{+0.05}_{-0.05}$&$-15.5\pm0.2$&$9.1$\\ 
b37&$700.0^{+41.0}_{-36.0}$&$3.88^{+0.26}_{-0.24}$&$-0.12^{+0.61}_{-0.2}$&$0.43^{+0.15}_{-0.1}$&$2.39^{+0.64}_{-0.71}$&$-4.99^{+0.18}_{-0.29}$&$-14.7\pm0.2$&$7.5$\\ 
\hline
\end{tabular}
}
\tablefoot{Same as Table \ref{tab:loo_a} but the models where performed using a uniform prior for log($g$) from 3 to 4\,dex.}
\label{tab:loo_b}
\end{table*}

\end{appendix}

\end{document}